\documentclass[useAMS, usenatbib]{mn2e}
\usepackage{color}
\usepackage{epsfig}
\usepackage{epstopdf}
\usepackage{amsmath, amssymb}
\usepackage{hyperref}
\usepackage{wasysym} 
\usepackage{subfigure}

\usepackage{setspace}  

\newlength{\plotwidth}
\newlength{\fullwidth}
\title[6dFGS: BAOs and the Local Hubble Constant]{The 6dF Galaxy Survey: Baryon Acoustic Oscillations and the Local Hubble Constant}
\author[Florian Beutler et al.]
{\parbox{\textwidth}{Florian Beutler$^{1}$\thanks{E-mail: \texttt{florian.beutler@icrar.org}},
Chris Blake$^2$, Matthew Colless$^3$, D. Heath Jones$^3$,\\
Lister Staveley-Smith$^{1}$,  Lachlan Campbell$^4$, Quentin Parker$^{3,5}$, Will Saunders$^3$, Fred Watson$^3$}\vspace{0.4cm}\\
\parbox{\textwidth}{
$^{1}$International Centre for Radio Astronomy Research (ICRAR), University of Western Australia, 35 Stirling Highway, Perth WA 6009, Australia\\
$^{2}$Centre for Astrophysics \& Supercomputing, Swinburne University of Technology, P.O. Box 218, Hawthorn, VIC 3122, Australia\\
$^{3}$Australian Astronomical Observatory, PO Box 296, Epping NSW 1710, Australia\\
$^{4}$Western Kentucky University, Bowling Green, USA\\
$^{5}$Macquarie University, Sydney, Australia}}

\begin{document}

\label{firstpage}

\maketitle

\begin{abstract}
We analyse the large-scale correlation function of the 6dF Galaxy Survey (6dFGS) and detect a Baryon Acoustic Oscillation (BAO) signal. The 6dFGS BAO detection allows us to constrain the distance-redshift relation at $z_{\rm eff} = 0.106$. We achieve a distance measure of $D_V(z_{\rm eff}) = 456\pm27\;$Mpc and a measurement of the distance ratio, $r_s(z_d)/D_V(z_{\rm eff}) = 0.336\pm0.015$ ($4.5\%$ precision), where $r_s(z_d)$ is the sound horizon at the drag epoch $z_d$. The low effective redshift of 6dFGS makes it a competitive and independent alternative to Cepheids and low-$z$ supernovae in constraining the Hubble constant. We find a Hubble constant of $H_0 = 67\pm3.2$\;km\;s$^{-1}$\;Mpc$^{-1}$ ($4.8\%$ precision) that depends only on the WMAP-7 calibration of the sound horizon and on the galaxy clustering in 6dFGS. Compared to earlier BAO studies at higher redshift, our analysis is less dependent on other cosmological parameters. The sensitivity to $H_0$ can be used to break the degeneracy between the dark energy equation of state parameter $w$ and $H_0$ in the CMB data. We determine that $w = -0.97\pm0.13$, using only WMAP-7 and BAO data from both 6dFGS and~\citet{Percival:2009xn}. 

We also discuss predictions for the large scale correlation function of two future wide-angle surveys: the WALLABY blind H{\sc I} survey (with the Australian SKA Pathfinder, ASKAP), and the proposed TAIPAN all-southern-sky optical galaxy survey with the UK Schmidt Telescope (UKST). We find that both surveys are very likely to yield detections of the BAO peak, making WALLABY the first radio galaxy survey to do so. We also predict that TAIPAN has the potential to constrain the Hubble constant with $3\%$ precision.
\end{abstract}


\section{Introduction}

The current standard cosmological model, $\Lambda$CDM, assumes that the initial fluctuations in the distribution of matter were seeded by quantum fluctuations pushed to cosmological scales by inflation. Directly after inflation, the universe is radiation dominated and the baryonic matter is ionised and coupled to radiation through Thomson scattering. The radiation pressure drives sound-waves originating from over-densities in the matter distribution~\citep{Peebles:1970ag,Sunyaev:1970eu,Bond:1987ub}. At the time of recombination ($z_* \approx 1090$) the photons decouple from the baryons and shortly after that (at the baryon drag epoch $z_d \approx 1020$) the sound wave stalls. Through this process each over-density of the original density perturbation field has evolved to become a centrally peaked perturbation surrounded by a spherical shell~\citep{Bashinsky:2000uh,Bashinsky:2002vx, Eisenstein:2006nj}. The radius of these shells is called the sound horizon $r_s$. Both over-dense regions attract baryons and dark matter and will be preferred regions of galaxy formation. This process can equivalently be described in Fourier space, where during the photon-baryon coupling phase, the amplitude of the baryon perturbations cannot grow and instead undergo harmonic motion leading to an oscillation pattern in the power spectrum.

After the time of recombination, the mean free path of photons increases and becomes larger than the Hubble distance. Hence from now on the radiation remains almost undisturbed, eventually becoming the Cosmic Microwave Background (CMB).

The CMB is a powerful probe of cosmology due to the good theoretical understanding of the physical processes described above. The size of the sound horizon depends (to first order) only on the sound speed in the early universe and the age of the Universe at recombination, both set by the physical matter and baryon densities, $\Omega_mh^2$ and $\Omega_bh^2$~\citep{Eisenstein:1997ik}. Hence, measuring the sound horizon in the CMB gives extremely accurate constraints on these quantities~\citep{Komatsu:2010fb}. Measurements of other cosmological parameters often show degeneracies in the CMB data alone~\citep{Efstathiou:1998xx}, especially in models with extra parameters beyond flat $\Lambda$CDM. Combining low redshift data with the CMB can break these degeneracies.

Within galaxy redshift surveys we can use the correlation function, $\xi$, to quantify the clustering on different scales. The sound horizon marks a preferred separation of galaxies and hence predicts a peak in the correlation function at the corresponding scale. The expected enhancement at $s = r_s$ is only $\Delta\xi \approx 10^{-3}b^2(1 + 2\beta/3 + \beta^2/5)$ in the galaxy correlation function, where $b$ is the galaxy bias compared to the matter correlation function and $\beta$ accounts for linear redshift space distortions. Since the signal appears at very large scales, it is necessary to probe a large volume of the universe to decrease sample variance, which dominates the error on these scales~\citep{Tegmark:1997rp,Goldberg:1997kv,Eisenstein:1998tu}.

Very interesting for cosmology is the idea of using the sound horizon scale as a standard ruler~\citep{Eisenstein:1998tu,Cooray:2001av, Seo:2003pu, Blake:2003rh}. A standard ruler is a feature whose absolute size is known. By measuring its apparent size, one can determine its distance from the observer. The BAO signal can be measured in the matter distribution at low redshift, with the CMB calibrating the absolute size, and hence the distance-redshift relation can be mapped (see e.g.~\cite{Bassett:2009mm} for a summary).

The Sloan Digital Sky Survey (SDSS; \citealt{York:2000gk}), and the 2dF Galaxy Redshift Survey (2dFGRS; \citealt{Colless:2001gk}) were the first redshift surveys which have directly detected the BAO signal. Recently the WiggleZ Dark Energy Survey has reported a BAO measurement at redshift $z=0.6$~\citep{Blake:2011wn}.

\cite{Eisenstein:2005su} were able to constrain the distance-redshift relation to $5\%$ accuracy at an effective redshift of $z_{\rm eff} = 0.35$ using an early data release of the SDSS-LRG sample containing $\approx 47\,000$ galaxies. Subsequent studies using the final SDSS-LRG sample and combining it with the SDSS-main and the 2dFGRS sample were able to improve on this measurement and constrain the distance-redshift relation at $z_{\rm eff} = 0.2$ and $z_{\rm eff} = 0.35$ with $3\%$ accuracy~\citep{Percival:2009xn}. Other studies of the same data found similar results using the correlation function $\xi(s)$ \citep{Martinez:2009,Gaztanaga:2008xz,Labini:2009ke,Sanchez:2009jq,Kazin:2009cj}, the power spectrum $P(k)$ \citep{Cole:2005sx,Tegmark:2006az,Huetsi:2006gu,Reid:2009xm}, the projected correlation function $w(r_p)$ of photometric redshift samples \citep{Padmanabhan:2006ku,Blake:2006kv} and a cluster sample based on the SDSS photometric data~\citep{Huetsi:2009zq}. Several years earlier a study by~\cite{Miller:2001cf} found first hints of the BAO feature in a combination of smaller datasets.

Low redshift distance measurements can directly measure the Hubble constant $H_0$ with a relatively weak dependence on other cosmological parameters such as the dark energy equation of state parameter $w$. The 6dF Galaxy Survey is the biggest galaxy survey in the local universe, covering almost half the sky. If 6dFGS could be used to constrain the redshift-distance relation through baryon acoustic oscillations, such a measurement could directly determine the Hubble constant, depending only on the calibration of the sound horizon through the matter and baryon density. The objective of the present paper is to measure the two-point correlation function on large scales for the 6dF Galaxy Survey and extract the BAO signal.

Many cosmological parameter studies add a prior on $H_0$ to help break degeneracies. The 6dFGS derivation of $H_0$ can provide an alternative source of that prior. The 6dFGS $H_0$-measurement can also be used as a consistency check of other low redshift distance calibrators such as Cepheid variables and Type Ia supernovae (through the so called distance ladder technique; see e.g.~\citealp{Freedman:2000cf,Riess:2011yx}). Compared to these more classical probes of the Hubble constant, the BAO analysis has an advantage of simplicity, depending only on $\Omega_mh^2$ and $\Omega_bh^2$ from the CMB and the sound horizon measurement in the correlation function, with small systematic uncertainties. 

Another motivation for our study is that the SDSS data after data release 3 (DR3) show more correlation on large scales than expected by $\Lambda$CDM and have no sign of a cross-over to negative $\xi$ up to $200h^{-1}\;$Mpc (the $\Lambda$CDM prediction is $140h^{-1}\;$Mpc)~\citep{Kazin:2009cj}. It could be that the LRG sample is a rather unusual realisation, and the additional power just reflects sample variance. It is interesting to test the location of the cross-over scale in another red galaxy sample at a different redshift.

This paper is organised as follows. In Section~\ref{sec:6dF} we introduce the 6dFGS survey and the $K$-band selected sub-sample used in this analysis. In Section~\ref{sec:analysis} we explain the technique we apply to derive the correlation function and summarise our error estimate, which is based on log-normal realisations. In Section~\ref{sec:model} we discuss the need for wide angle corrections and several linear and non-linear effects which influence our measurement. Based on this discussion we introduce our correlation function model. In Section~\ref{sec:DV} we fit the data and derive the distance estimate $D_V(z_{\rm eff})$. In Section~\ref{sec:results} we derive the Hubble constant and constraints on dark energy. In Section~\ref{sec:sig} we discuss the significance of the BAO detection of 6dFGS. In Section~\ref{sec:future} we give a short overview of future all-sky surveys and their power to measure the Hubble constant. We conclude and summarise our results in Section~\ref{sec:conclusion}.\\

Throughout the paper, we use $r$ to denote real space separations and $s$ to denote separations in redshift space. Our fiducial model assume a flat universe with $\Omega^{\rm fid}_m = 0.27$, $w^{\rm fid} = -1$ and $\Omega_k^{\rm fid} = 0$. The Hubble constant is set to $H_0 = 100h\;$km s$^{-1}$Mpc$^{-1}$, with our fiducial model using $h^{\rm fid} = 0.7$.

\section{The 6dF galaxy survey}
\label{sec:6dF}

\subsection{Targets and Selection Function}
\label{sec:selectionfn}

The galaxies used in this analysis were selected to $K \leq 12.9$ from the 2MASS Extended Source Catalog~\citep[2MASS XSC;][]{Jarrett:2000me} and combined with redshift data from the 6dF Galaxy Survey \citep[6dFGS;][]{Jones:2009yz}. The 6dF Galaxy Survey is a combined redshift and peculiar velocity survey covering nearly the entire southern sky with $|b| < 10^\circ$. It was undertaken with the Six-Degree Field (6dF) multi-fibre instrument on the UK Schmidt Telescope from 2001 to 2006. The median redshift of the survey is $z = 0.052$ and the $25\%:50\%:75\%$ percentile completeness values are $0.61:0.79:0.92$. Papers by~\citet{Jones:2004zy, Jones:2006xy, Jones:2009yz} describe 6dFGS in full detail, including comparisons between 6dFGS, 2dFGRS and SDSS.

Galaxies were excluded from our sample if they resided in sky regions with completeness lower than 60 percent. After applying these cuts our sample contains $75\,117$ galaxies. The selection function was derived by scaling the survey completeness as a function of magnitude to match the integrated on-sky completeness, using mean galaxy counts. This method is the same adopted by~\citet{Colless:2001gk} for 2dFGRS and is explained in~\citet{Jones:2006xy} in detail. The redshift of each object was checked visually and care was taken to exclude foreground Galactic sources. The derived completeness function was used in the real galaxy catalogue to weight each galaxy by its inverse completeness. The completeness function was also applied to the mock galaxy catalogues to mimic the selection characteristics of the survey. Jones et al. (in preparation) describe the derivation of the 6dFGS selection function, and interested readers are referred to this paper for a more comprehensive treatment.

\subsection{Survey volume}

\begin{figure}
\begin{center}
\epsfig{file=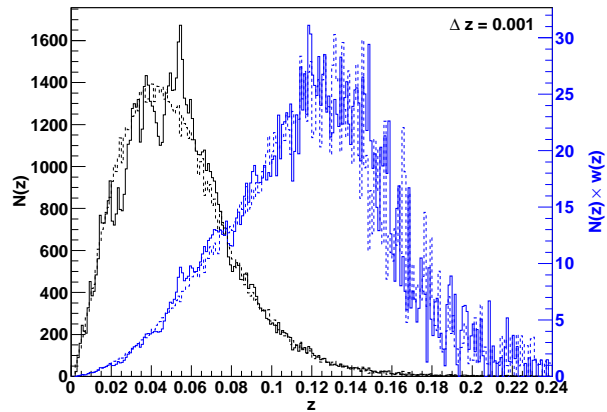,width=8cm}
\caption{Redshift distribution of the data (black solid line) and the random catalogue (black dashed line). The weighted 
distribution (using weights from eq.~\ref{eq:weight}) is shifted to higher redshift and has increased shot noise 
but a smaller error due to sample variance (blue solid and dashed lines).}
\label{fig:wred}
\end{center}
\end{figure}

We calculated the effective volume of the survey using the estimate of~\citet{Tegmark:1997rp}
\begin{equation}
V_{\rm eff} = \int d^3\vec{x} \left[\frac{n(\vec{x})P_0}{1 + n(\vec{x})P_0}\right]^2
\label{eq:veff}
\end{equation}
where $n(\vec{x})$ is the mean galaxy density at position $\vec{x}$, determined from the data, and $P_0$ is the characteristic power spectrum amplitude of the BAO signal. The parameter $P_0$ is crucial for the weighting scheme introduced later. We find that the value of $P_0 = 40\,000h^{-3}\;$Mpc$^3$ (corresponding to the value of the galaxy power spectrum at $k \approx 0.06h\;$Mpc$^{-1}$ in 6dFGS) minimises the error of the correlation function near the BAO peak.

Using $P_0 = 40\,000h^{-3}\;$Mpc$^3$ yields an effective volume of $0.08h^{-3}\;$Gpc$^3$, while using instead $P_0 = 10\,000h^{-3}\;$Mpc$^3$ (corresponding to $k \approx 0.15h\;$Mpc$^{-1}$) gives an effective volume of $0.045h^{-3}\;$Gpc$^3$.

The volume of the 6dF Galaxy Survey is approximately as large as the volume covered by the 2dF Galaxy Redshift Survey, with a sample density similar to SDSS-DR7~\citep{Abazajian:2008wr}. \citet{Percival:2009xn} reported successful BAO detections in several samples obtained from a combination of SDSS DR7, SDSS-LRG and 2dFGRS with effective volumes in the range $0.15 - 0.45h^{-3}\;$Gpc$^3$ (using $P_0 = 10\,000h^{-3}\;$Mpc$^3$), while the original detection by~\citet{Eisenstein:2005su} used a sample with $V_{\rm eff} = 0.38h^{-3}\;$Gpc$^3$ (using $P_0 = 40\,000h^{-3}\;$Mpc$^3$).

\section{Clustering measurement}
\label{sec:analysis}

We focus our analysis on the two-point correlation function. In the following sub-sections we introduce the technique used to estimate the correlation function and outline the method of log-normal realisations, which we employed to derive a covariance matrix for our measurement.

\subsection{Random catalogues}

To calculate the correlation function we need a random sample of galaxies which follows the same angular and redshift selection function as the 6dFGS sample. We base our random catalogue generation on the 6dFGS luminosity function of Jones et al. (in preparation), where we use random numbers to pick volume-weighted redshifts and luminosity function-weighted absolute magnitudes. We then test whether the redshift-magnitude combination falls within the 6dFGS $K$-band faint and bright apparent magnitude limits ($8.75 \leq K \leq 12.9$).

Figure~\ref{fig:wred} shows the redshift distribution of the 6dFGS $K$-selected sample (black solid line) compared to a random catalogue with the same number of galaxies (black dashed line). The random catalogue is a good description of the 6dFGS redshift distribution in both the weighted and unweighted case.

\subsection{The correlation function}
\label{sec:cor}

\begin{figure}
\begin{center}
\epsfig{file=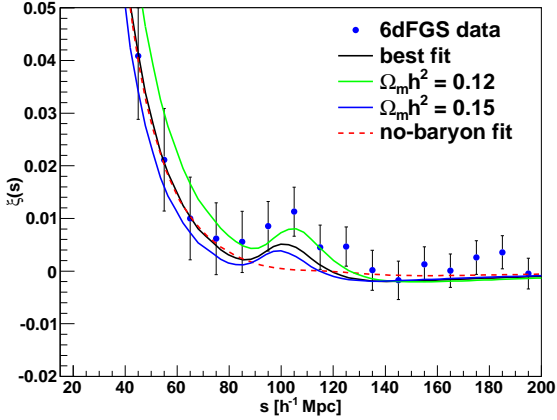,width=8cm}
\caption{The large scale correlation function of 6dFGS. The best fit model is shown by the black line with the best fit value of $\Omega_mh^2 = 0.138\pm0.020$. Models with different $\Omega_mh^2$ are shown by the green line ($\Omega_mh^2 = 0.12$) and the blue line ($\Omega_mh^2 = 0.15$). The red dashed line is a linear CDM model with $\Omega_bh^2 = 0$ (and $\Omega_mh^2 = 0.1$), while all other models use the WMAP-7 best fit value of $\Omega_bh^2 = 0.02227$~{\protect \citep{Komatsu:2010fb}}. The significance of the BAO detection in the black line relative to the red dashed line is $2.4\sigma$ (see Section~\ref{sec:sig}). The error-bars at the data points are the diagonal elements of the covariance matrix derived using
log-normal mock catalogues.}
\label{fig:bao}
\end{center}
\end{figure}

We turn the measured redshift into co-moving distance via
\begin{equation}
D_C(z) = \frac{c}{H_0}\int^z_0\frac{dz'}{E(z')}
\end{equation} 
with
\begin{align}
E(z) = \big[&\Omega^{\rm fid}_m(1+z)^3 + \Omega^{\rm fid}_k(1+z)^2\cr
             &+ \Omega^{\rm fid}_{\Lambda}(1+z)^{3(1+w^{\rm fid})})\big]^{1/2},
\end{align}
where the curvature $\Omega^{\rm fid}_k$ is set to zero, the dark energy density is given by $\Omega^{\rm fid}_{\Lambda} = 1-\Omega^{\rm fid}_m$ and the equation of state for dark energy is $w^{\rm fid} = -1$. Because of the very low redshift of 6dFGS, our data are not very sensitive to $\Omega_k$, $w$ or any other higher dimensional parameter which influences the expansion history of the universe. We will discuss this further in Section~\ref{sec:AandR}.

Now we measure the separation between all galaxy pairs in our survey and count the number of such pairs in each separation bin. We do this for the 6dFGS data catalogue, a random catalogue with the same selection function and a combination of data-random pairs. We call the pair-separation distributions obtained from this analysis $DD(s), RR(s)$ and $DR(s)$, respectively. The binning is chosen to be from $10h^{-1}$\;Mpc up to $190h^{-1}$\;Mpc, in $10h^{-1}$\;Mpc steps. In the analysis we used $30$ random catalogues with the same size as the data catalogue. The redshift correlation function itself is given by~\citet{Landy:1993yu}:
\begin{equation}
\xi'_{\rm data}(s) = 1 + \frac{DD(s)}{RR(s)} \left(\frac{n_r}{n_d} \right)^2 - 2\frac{DR(s)}{RR(s)} \left(\frac{n_r}{n_d} \right),
\label{eq:LS2}
\end{equation}
where the ratio $n_r/n_d$ is given by
\begin{equation}
\frac{n_r}{n_d} = \frac{\sum^{N_r}_iw_i(\vec{x})}{\sum^{N_d}_jw_j(\vec{x})}
\end{equation}
and the sums go over all random ($N_r$) and data ($N_d$) galaxies. We use the inverse density weighting of~\citet{Feldman:1993ky}:
\begin{equation}
w_i(\vec{x}) = \frac{C_i}{1 + n(\vec{x})P_0},
\label{eq:weight}
\end{equation}
with $P_0 = 40\,000h^{3}$\;Mpc$^{-3}$ and $C_i$ being the inverse completeness weighting for 6dFGS (see Section~\ref{sec:selectionfn} and Jones et al., in preparation). This weighting is designed to minimise the error on the BAO measurement, and since our sample is strongly limited by sample variance on large scales this weighting results in a significant improvement to the analysis. The effect of the weighting on the redshift distribution is illustrated in Figure~\ref{fig:wred}.

Other authors have used the so called $J_3$-weighting which optimises the error over all scales by weighting each scale differently~\citep[e.g.][]{Efstathiou:1988, Loveday:1994dx}. In a magnitude limited sample there is a correlation between luminosity and redshift, which establishes a correlation between bias and redshift~\citep{Zehavi:2004zn}. A scale-dependent weighting would imply a different effective redshift for each scale, causing a scale dependent bias. 

Finally we considered a luminosity dependent weighting as suggested by~\citet{Percival:2003pi}. However the same authors found that explicitly accounting for the luminosity-redshift relation has a negligible effect for 2dFGRS. We found that the effect to the 6dFGS correlation function is $\ll 1\sigma$ for all bins. Hence the static weighting of eq.~\ref{eq:weight} is sufficient for our dataset.

We also include an integral constraint correction in the form of 
\begin{equation}
\xi_{\rm data}(s) = \xi'_{\rm data}(s) + ic,
\end{equation}
where $ic$ is defined as
\begin{equation}
ic = \frac{\sum_s RR(s)\xi_{\rm model}(s)}{\sum_s RR(s)}.
\end{equation}
The function $RR(s)$ is calculated from our mock catalogue and $\xi_{\rm model}(s)$ is a correlation function model. Since $ic$ depends on the model of the correlation function we have to re-calculate it at each step during the fitting procedure. However we note that $ic$ has no significant impact to the final result.

Figure~\ref{fig:bao} shows the correlation function of 6dFGS at large scales. The BAO peak at $\approx 105h^{-1}\;$Mpc is clearly visible. The plot includes model predictions of different cosmological parameter sets. We will discuss these models in Section~\ref{sec:DV2}. 

\subsection{Log-normal error estimate}
\label{sec:log}

\begin{figure}
\begin{center}
\epsfig{file=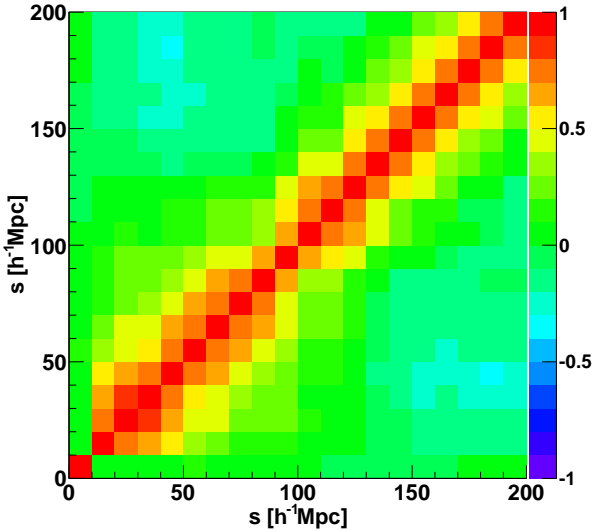,width=8cm}
\caption{Correlation matrix derived from a covariance matrix calculated from $200$ log-normal realisations. }
\label{fig:matrix_bao}
\end{center}
\end{figure}

To obtain reliable error-bars for the correlation function we use log-normal realisations~\citep{Coles:1991if, Cole:2005sx, Kitaura:2009jc}. In what follows we summarise the main steps, but refer the interested reader to Appendix~\ref{ap:log} 
in which we give a detailed explanation of how we generate the log-normal mock catalogues. 
In Appendix~\ref{ap:jk_comp} we compare the log-normal errors with jack-knife estimates.\\

Log-normal realisations of a galaxy survey are usually obtained by deriving a density field from a model power spectrum, $P(k)$, assuming Gaussian fluctuations. This density field is then Poisson sampled, taking into account the window function and the total number of galaxies. The assumption that the input power spectrum has Gaussian fluctuations can only be used in a model for a density field with over-densities $\ll 1$. As soon as we start to deal with finite rms fluctuations, the Gaussian model assigns a non-zero probability to regions of negative density. A log-normal random field $LN(\vec{x})$, can avoid this unphysical behaviour. It is obtained from a Gaussian field $G(\vec{x})$ by
\begin{equation}
LN(\vec{x}) = \exp[G(\vec{x})]
\end{equation}
which is  positive-definite but approaches $1+G(\vec{x})$ whenever the perturbations are small (e.g. at large scales). Calculating the power spectrum of a Poisson sampled density field with such a distribution will reproduce the input power spectrum convolved with the window function. As an input power spectrum for the log-normal field we use
\begin{equation}
P_{\rm nl}(k) = A P_{\rm lin}(k)\exp[-(k/k_*)^2]
\end{equation}
where $A = b^2(1 + 2\beta/3 + \beta^2/5)$ accounts for the linear bias and the linear redshift space distortions. $P_{\rm lin}(k)$ is a linear model power spectrum in real space obtained from CAMB~\citep{Lewis:1999bs} and $P_{\rm nl}(k)$ is the non-linear power spectrum in redshift space. Comparing the model above with the 6dFGS data gives $A = 4$. The damping parameter $k_*$ is set to $k_* = 0.33h$\;Mpc$^{-1}$, as found in 6dFGS (see fitting results later). How well this input model matches the 6dFGS data can be seen in Figure~\ref{fig:log_6df}.

We produce $200$ such realisations and calculate the correlation function for each of them, deriving a covariance matrix
\begin{equation}
C_{ij} = \sum^N_{n=1}\frac{\left[\xi_n(s_i) - \overline{\xi}(s_i)\right]\left[\xi_n(s_j) - \overline{\xi}(s_j)\right]}{N-1} .
\end{equation}
Here, $\xi_n(s_i)$ is the correlation function estimate at separation $s_i$ and the sum goes over all $N$ log-normal realisations. The mean value is defined as
\begin{equation}
\overline{\xi}(r_i) = \frac{1}{N}\sum^N_{n=1}\xi_n(s_i) .
\end{equation}
The case $i = j$ gives the error (ignoring correlations between bins, $\sigma_i^2 = C_{ii}$). In the following we will use this uncertainty in all diagrams, while the fitting procedures use the full covariance matrix.

The distribution of recovered correlation functions includes the effects of sample variance and shot noise. Non-linearities are also approximately included since the distribution of over-densities is skewed.

In Figure~\ref{fig:matrix_bao} we show the log-normal correlation matrix $r_{ij}$ calculated from the covariance matrix. The correlation matrix is defined as
\begin{equation}
r_{ij} = \frac{C_{ij}}{\sqrt{C_{ii}C_{jj}}},
\end{equation}
where $C$ is the covariance matrix (for a comparison to jack-knife errors see appendix~\ref{ap:jk_comp}).

\section{Modelling the BAO signal}
\label{sec:model}

In this section we will discuss wide-angle effects and non-linearities. We also introduce a model for the large scale correlation function, which we later use to fit our data.

\subsection{Wide angle formalism}

The model of linear redshift space distortions introduced by~\citet{Kaiser:1987qv} is based on the plane parallel approximation. Earlier surveys such as SDSS and 2dFGRS are at sufficiently high redshift that the maximum opening angle between a galaxy pair remains small enough to ensure the plane parallel approximation is valid. However,
the 6dF Galaxy Survey has a maximum opening angle of $180^{\circ}$ and a lower mean redshift of $z\approx 0.1$ (for our weighted sample) and so it is necessary to test the validity of  the plane parallel approximation. 
The wide angle description of redshift space distortions has been laid out in several papers~\citep{Szalay:1997cc,Szapudi:2004gh,Matsubara:2004fr,Papai:2008bd,Raccanelli:2010hk}, which we summarise in Appendix~\ref{ap:wide}.

We find that the wide-angle corrections have only a very minor effect on our sample. For our fiducial model we found a correction of $\Delta\xi = 4\cdot10^{-4} $ in amplitude at $s = 100h^{-1}$\;Mpc and $\Delta\xi = 4.5\cdot10^{-4}$ at $s = 200h^{-1}$\;Mpc, (Figure~\ref{fig:wide2} in the appendix). This is much smaller than the error bars on these scales. Despite the small size of the effect, we nevertheless include all first order correction terms in our correlation function model. It is important to note that wide angle corrections affect the correlation function amplitude only and do not cause any shift in the BAO scale. The effect of the wide-angle correction on the unweighted sample is much greater and is already noticeable on scales of $20h^{-1}$\;Mpc. Weighting to higher redshifts mitigates the effect because it reduces the average opening angle between galaxy pairs, by giving less weight to wide angle pairs (on average).

\subsection{Non-linear effects}
\label{sec:nlin}

There are a number of non-linear effects which can potentially influence a measurement of the BAO signal.
These include scale-dependent bias, the non-linear growth of structure on smaller scales, and redshift
space distortions. We discuss each of these in the context of our 6dFGS sample.

As the universe evolves, the acoustic signature in the correlation function is broadened by non-linear gravitational structure formation. Equivalently we can say that the higher harmonics in the power spectrum, which represent smaller scales, are erased~\citep{Eisenstein:2006nj}.

The early universe physics, which we discussed briefly in the introduction, is well understood and several authors have produced software packages (e.g. CMBFAST and CAMB) and published fitting functions \citep[e.g ][]{Eisenstein:1997ik} to make predictions for the correlation function and power spectrum using thermodynamical models of the early universe. These models already include the basic linear physics most relevant for the BAO peak. In our analysis we use the CAMB software package~\citep{Lewis:1999bs}. The non-linear evolution of the power spectrum in CAMB is calculated using the halofit code~\citep{Smith:2002dz}. This code is calibrated by $n$-body simulations and can describe non-linear effects in the shape of the matter power spectrum for pure CDM models to an accuracy of around $5-10\%$~\citep{Heitmann:2008eq}. However, it has previously been shown that this non-linear model is a poor description of the non-linear effects around the BAO peak~\citep{Crocce:2007dt}. We therefore decided to use the linear model output from CAMB and incorporate the non-linear effects separately.

All non-linear effects influencing the correlation function can be approximated by a convolution with a Gaussian damping factor $\exp[-(rk_*/2)^2]$~\citep{Eisenstein:2006nk, Eisenstein:2006nj}, where $k_*$ is the damping scale. We will use this factor in our correlation function model introduced in the next section. The convolution with a Gaussian causes a shift of the peak position to larger scales, since the correlation function is not symmetric around the peak. However this shift is usually very small.

All of the non-linear processes discussed so far are not at the fundamental scale of $105h^{-1}$\;Mpc but are instead at the cluster-formation scale of up to $10h^{-1}$Mpc. The scale of $105h^{-1}$\;Mpc is far larger than any known non-linear effect in cosmology. This has led some authors to the conclusion that the peak will not be shifted significantly, but rather only blurred out. For example,~\citet{Eisenstein:2006nj} have argued that any systematic shift of the acoustic scale in real space must be small ($\apprle 0.5\%$), even at $z = 0$. 

However, several authors report possible shifts of up to $1\%$~\citep{Guzik:2006bu,Smith:2007gi,Smith:2006ne,Angulo:2007fw}. \citet{Crocce:2007dt} used re-normalised perturbation theory (RPT) and found percent-level shifts in the BAO peak. In addition to non-linear evolution, they found that mode-coupling generates additional oscillations in the power spectrum, which are out of phase with the BAO oscillations predicted by linear theory. This leads to shifts in the scale of oscillation nodes with respect to a smooth spectrum. In real space this corresponds to a peak shift towards smaller scales.
Based on their results,~\citet{Crocce:2007dt} propose a model to be used for the correlation function analysis at large scales. We will introduce this model in the next section.

\subsection{Large-scale correlation function}
\label{sec:lss}

To model the correlation function on large scales, we follow~\citet{Crocce:2007dt} and~\citet{Sanchez:2008iw} and adopt the following parametrisation\footnote{note that $r=s$, the different letters just specify whether the function is evaluated in redshift space or real space.}: 
\begin{equation}
\xi'_{\rm model}(s) = B(s)b^2\left[\xi(s) * G(r) + \xi^1_1(r)\frac{\partial \xi(s)}{\partial s}\right].
\label{eq:scocci1}
\end{equation}
Here, we decouple the scale dependency of the bias $B(s)$ and the linear bias $b$. $G(r)$ is a Gaussian damping term, accounting for non-linear suppression of the BAO signal. $\xi(s)$ is the linear correlation function (including wide angle description of redshift space distortions; eq.~\ref{eq:mom} in the appendix). The second term in eq~\ref{eq:scocci1} accounts for the mode-coupling of different Fourier modes. It contains $\partial \xi(s)/\partial s$, which is the first derivative of the redshift space correlation function, and $\xi^1_1(r)$, which is defined as 
\begin{equation}
\xi^1_1(r) = \frac{1}{2\pi^2}\int^{\infty}_0 dk\;kP_{\rm lin}(k)j_1(rk) ,
\end{equation}
with $j_1(x)$ being the spherical Bessel function of order $1$. 
\citet{Sanchez:2008iw} used an additional parameter $A_{\rm MC}$ which multiplies the mode coupling term in equation~\ref{eq:scocci1}. We found that our data is not good enough to constrain this parameter, and hence adopted $A_{\rm MC} = 1$ as in the original model by~\citet{Crocce:2007dt}.

In practice we generate linear model power spectra $P_{\rm lin}(k)$ from CAMB and convert them into a correlation function using a Hankel transform
\begin{equation}
\xi(r) = \frac{1}{2\pi^2}\int^{\infty}_0 dk\;k^2P_{\rm lin}(k)j_0(rk),
\end{equation}
where $j_0(x) = \sin(x)/x$ is the spherical Bessel function of order $0$.

The $*$-symbol in eq.~\ref{eq:scocci1} is only equivalent to a convolution in the case of a 3D correlation function, where we have the Fourier theorem relating the 3D power spectrum to the correlation function. In case of the spherically averaged quantities this is not true. Hence, the $*$-symbol in our equation stands for the multiplication of the power spectrum with $\tilde{G}(k)$ before transforming it into a correlation function. $\tilde{G}(k)$ is defined as 
\begin{equation}
\tilde{G}(k) = \exp\left[-(k/k_*)^2\right],
\label{eq:damp}
\end{equation}
with the property
\begin{equation}
\tilde{G}(k) \rightarrow 0 \;\text{ as }\; k \rightarrow \infty.
\end{equation}
The damping scale $k_*$ can be calculated from linear theory~\citep{Crocce:2005xy,Matsubara:2007wj} by
\begin{equation}
k_* = \left[\frac{1}{6\pi^2}\int^{\infty}_0 dk\;P_{\rm lin}(k)\right]^{-1/2},
\label{eq:sigv}
\end{equation}
where $P_{\rm lin}(k)$ is again the linear power spectrum. $\Lambda$CDM predicts a value of $k_* \simeq 0.17h\;$Mpc$^{-1}$. However, we will include $k_*$ as a free fitting parameter.

The scale dependance of the 6dFGS bias, $B(s)$, is derived from the GiggleZ simulation (Poole et al., in preparation); a dark matter simulation containing $2160^3$ particles in a $1h^{-1}\;$Gpc box.  We rank-order the halos of this simulation by $V_{\rm max}$ and choose a contiguous set of $250\,000$ of them, selected to have the same clustering amplitude of 6dFGS as quantified by the separation scale $r_0$, where $\xi(r_0) = 1$. In the case of 6dFGS we found $r_0 = 9.3h^{-1}\;$Mpc. Using the redshift space correlation function of these halos and of a randomly subsampled set of $\sim {10^6}$ dark matter particles, we obtain
\begin{equation}
B(s) = 1 + \left(s/0.474h^{-1}\text{Mpc}\right)^{-1.332},
\end{equation}
which describes a $1.7\%$ correction of the correlation function amplitude at separation scales of $10h^{-1}\;$Mpc. To derive this function, the GiggleZ correlation function (snapshot $z = 0$) has been fitted down to $6h^{-1}\;$Mpc, well below the smallest scales we are interested in.

\section{Extracting the BAO signal}
\label{sec:DV}

\begin{table}
\begin{center}
\caption{This table contains all parameter constraints from 6dFGS obtained in this paper. The priors used to derive these parameters are listed in square brackets. All parameters assume $\Omega_bh^2 = 0.02227$ and in cases where a prior on $\Omega_mh^2$ is used, we adopt the WMAP-7 Markov chain probability distribution~{\protect \citep{Komatsu:2010fb}}. $A(z_{\rm eff})$ is the acoustic parameter defined by~{\protect \citet{Eisenstein:2005su}} (see equation~\ref{eq:A} in the text) and $R(z_{\rm eff})$ is the distance ratio of the 6dFGS BAO measurement to the last-scattering surface. The most sensible value for cosmological parameter constraints is $r_s(z_d)/D_V(z_{\rm eff})$, since this measurement is uncorrelated with $\Omega_mh^2$. The effective redshift of 6dFGS is $z_{\rm eff} = 0.106$ and the fitting range is from $10-190h^{-1}$\;Mpc.}
	\begin{tabular}{rll}
		\hline
  		\multicolumn{3}{c}{Summary of parameter constraints from 6dFGS}\\
		\hline
		$\Omega_mh^2$ & $0.138\pm 0.020$ ($14.5\%$)& \\
     		$D_V(z_{\rm eff})$ & $456\pm27\;$Mpc ($5.9\%$)& \\
     		$D_V(z_{\rm eff})$ & $459\pm18\;$Mpc ($3.9\%$)& $[\Omega_mh^2$ prior]\\
		$\mathbf{r_s(z_d)/D_V(z_{\rm \textbf{eff}})}$ & $\mathbf{0.336\pm0.015}$ ($\mathbf{4.5\%}$) & \\
		$R(z_{\rm eff})$ & $0.0324\pm 0.0015$ ($4.6\%$) & \\
		$A(z_{\rm eff})$ & $0.526\pm0.028$ ($5.3\%$) & \\
		\hline
		$\Omega_m$ & $0.296\pm0.028$ ($9.5\%$) & $[\Omega_mh^2$ prior]\\
		$\mathbf{H_0}$ & $\mathbf{67\pm3.2}$ $\mathbf{(4.8\%)}$ & $\mathbf{[\Omega_mh^2}$ \textbf{prior]}\\
		\hline
	  \end{tabular}
\label{tab:para}
\end{center}
\end{table}

In this section we fit the model correlation function developed in the previous section to our data. 
Such a fit can be used to derive the distance scale $D_V(z_{\rm eff})$ at the effective redshift of the survey.

\subsection{Fitting preparation}
\label{sec:fitprep}

The effective redshift of our sample is determined by
\begin{equation}
z_{\rm eff} = \sum^{N_b}_{i}\sum^{N_b}_{j}\frac{w_iw_j}{2N_b^2}(z_i + z_j),
\end{equation}
where $N_{b}$ is the number of galaxies in a particular separation bin and $w_i$ and $w_j$ are the weights for those galaxies 
from eq.~\ref{eq:weight}. We choose $z_{\rm eff}$ from bin $10$ which has the limits $100h^{-1}$\;Mpc and $110h^{-1}$\;Mpc and which gave $z_{\rm eff} = 0.106$. Other bins show values very similar to this, with a standard deviation of $\pm0.001$. The final result does not depend on a very precise determination of $z_{\rm eff}$, since we are not constraining a distance to the mean redshift, but a distance ratio (see equation~\ref{eq:alpha}, later). In fact, if the fiducial model is correct, the result is completely independent of $z_{\rm eff}$. Only if there is a $z$-dependent deviation from the fiducial model do we need $z_{\rm eff}$ to quantify this deviation at a specific redshift.\\

Along the line-of-sight, the BAO signal directly constrains the Hubble constant $H(z)$ at redshift $z$. When measured in a redshift shell, it constrains the angular diameter distance $D_A(z)$~\citep{Matsubara:2004fr}. In order to separately measure $D_A(z)$ and $H(z)$ we require a BAO detection in the 2D correlation function, where it will appear as a ring at around $105 h^{-1}$\;Mpc. Extremely large volumes are necessary for such a measurement. While there are studies that report a successful (but very low signal-to-noise) detection in the 2D correlation function using the SDSS-LRG data~\citep[e.g. ][but see also~\citealt{Kazin:2010nd}]{Gaztanaga:2008xz,Chuang:2011fy}, our sample does not allow this kind of analysis. Hence we restrict ourself to the 1D correlation function, where we measure a combination of $D_A(z)$ and $H(z)$. What we actually measure is a superposition of two angular measurements (R.A. and Dec.) and one line-of-sight measurement (redshift). To account for this mixture of measurements it is common to report the BAO distance constraints as~\citep{Eisenstein:2005su, Padmanabhan:2008ag}
\begin{equation}
D_V(z) = \left[(1+z)^2D_A^2(z)\frac{cz}{H_0E(z)}\right]^{1/3},
\label{eq:dv}
\end{equation}
where $D_A$ is the angular distance, which in the case of $\Omega_k = 0$ is given by $D_A(z) = D_C(z)/(1+z)$.

To derive model power spectra from CAMB we have to specify a complete cosmological model, which in the case of the simplest $\Lambda$CDM model ($\Omega_k = 0, w = -1$), is specified by six parameters: $\omega_{c}, \omega_{b}, n_s, \tau$, $A_s$ and $h$. These parameters are: the physical cold dark matter and baryon density, ($\omega_{c} = \Omega_ch^2,\; \omega_{b} = \Omega_bh^2$), the scalar spectral index, ($n_s$), the optical depth at recombination, ($\tau$), the scalar amplitude of the CMB temperature fluctuation, ($A_s$), and the Hubble constant in units of $100\;$km\;s$^{-1}$Mpc$^{-1}$ ($h$).

Our fit uses the parameter values from WMAP-7~\citep{Komatsu:2010fb}:  $\Omega_bh^2 = 0.02227$, $\tau = 0.085$ and $n_s = 0.966$ (maximum likelihood values). The scalar amplitude $A_s$ is set so that it results in $\sigma_8 = 0.8$, which depends on $\Omega_mh^2$. However $\sigma_8$ is degenerated with the bias parameter $b$ which is a free parameter in our fit. Furthermore, $h$ is set to $0.7$ in the fiducial model, but can vary freely in our fit through a scale distortion parameter $\alpha$, which enters the model as
\begin{equation}
\xi_{\rm model}(s) = \xi'_{\rm model}(\alpha s).
\end{equation}
This parameter accounts for deviations from the fiducial cosmological model, which we use to derive distances from the measured redshift. It is defined as~\citep{Eisenstein:2005su, Padmanabhan:2008ag}
\begin{equation}
\alpha = \frac{D_V(z_{\rm eff})}{D_V^{\rm fid}(z_{\rm eff})}.
\label{eq:alpha}
\end{equation}
The parameter $\alpha$ enables us to fit the correlation function derived with the fiducial model, without the need to re-calculate the correlation function for every new cosmological parameter set. 

At low redshift we can approximate $H(z) \approx H_0$, which results in
\begin{equation}
\alpha \approx \frac{H^{\rm fid}_0}{H_0} .
\label{eq:H0}
\end{equation}
Compared to the correct equation~\ref{eq:alpha} this approximation has an error of about $3\%$ at redshift $z = 0.1$ for our fiducial model. Since this is a significant systematic bias, we do not use this approximation at any point in our analysis.

\subsection{Extracting $D_V(z_{\rm eff})$ and $r_s(z_d)/D_V(z_{\rm eff})$}
\label{sec:DV2}

\begin{figure}
\begin{center}
\epsfig{file=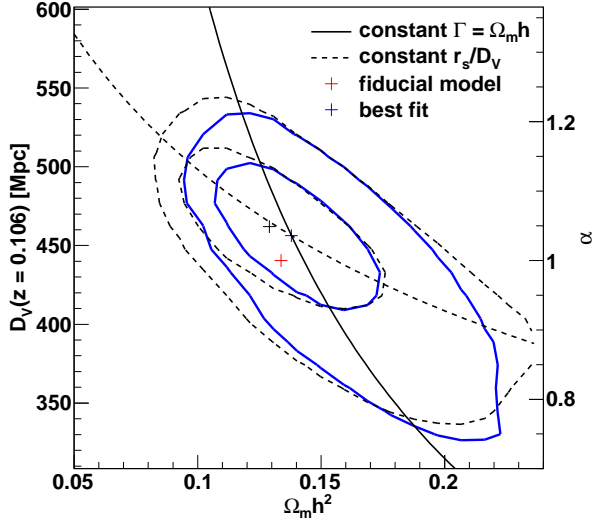,width=8cm}
\caption{Likelihood contours of the distance $D_V(z_{\rm eff})$ against $\Omega_mh^2$. The corresponding values of $\alpha$ are given on the right-hand axis. The contours show $1$ and  $2\sigma$ errors for both a full fit (blue solid contours) and a fit over $20-190h^{-1}$\;Mpc (black dashed contours) excluding the first data point. The black cross marks the best fitting values corresponding to the dashed black contours with $(D_V,\Omega_mh^2) = (462,0.129)$, while the blue cross marks the best fitting values for the blue contours. The black solid curve corresponds to a constant $\Omega_m h^2D_V(z_{\rm eff})$ ($D_V \sim h^{-1}$), while the dashed line corresponds to a constant angular size of the sound horizon, as described in the text.}
\label{fig:chi2}
\end{center}
\end{figure}

\label{sec:fit}
\begin{figure}
\begin{center}
\epsfig{file=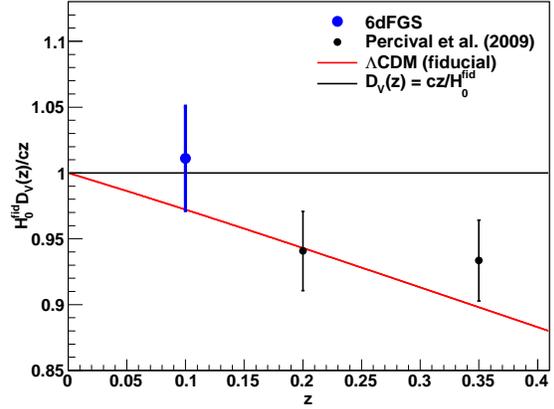,width=8cm}
\caption{The distance measurement $D_V(z)$ relative to a low redshift approximation. The points show 6dFGS data and those of~\citet{Percival:2009xn}.}
\label{fig:plot5_sn}
\end{center}
\end{figure}

Using the model introduced above we performed fits to $18$ data points between $10h^{-1}\;$Mpc and $190h^{-1}\;$Mpc. We excluded the data below $10h^{-1}\;$Mpc, since our model for non-linearities is not good enough to capture the effects on such scales. The upper limit is chosen to be well above the BAO scale, although the constraining contribution of the bins above $130h^{-1}\;$Mpc is very small. Our final model has $4$ free parameters:  $\Omega_mh^2$, $b$, $\alpha$ and $k_*$.

The best fit corresponds to a minimum $\chi^2$ of $15.7$ with $14$ degrees of freedom ($18$ data-points and $4$ free parameters). The best fitting model is included in Figure~\ref{fig:bao} (black line). The parameter values are $\Omega_m h^2 = 0.138\pm 0.020$, $b = 1.81\pm0.13$ and $\alpha = 1.036\pm0.062$, where the errors are derived for each parameter by marginalising over all other parameters. For $k_*$ we can give a lower limit of $k_* = 0.19h\;$Mpc$^{-1}$ (with $95\%$ confidence level).

We can use eq.~\ref{eq:alpha} to turn the measurement of $\alpha$ into a measurement of the distance to the effective redshift $D_V(z_{\rm eff}) = \alpha D_V^{\rm fid}(z_{\rm eff}) = 456\pm27\;$Mpc, with a precision of $5.9\%$. Our fiducial model gives $D_V^{\rm fid}(z_{\rm eff}) = 440.5\;$Mpc, where we have followed the distance definitions of~\citet{Wright:2006up} throughout. For each fit we derive the parameter $\beta = \Omega_m(z)^{0.545}/b$, which we need to calculate the wide angle corrections for the correlation function. 

The maximum likelihood distribution of $k_*$ seems to prefer smaller values than predicted by $\Lambda$CDM, although we are not able to constrain this parameter very well. This is connected to the high significance of the BAO peak in the 6dFGS data (see Section~\ref{sec:sig}). A smaller value of $k_*$ damps the BAO peak and weakens the distance constraint. For comparison we also performed a fit fixing $k_*$ to the $\Lambda$CDM prediction of $k_* \simeq 0.17h\;$Mpc$^{-1}$. We found that the error on the distance $D_V(z_{\rm eff})$ increases from $5.9\%$ to $8\%$. However since the data do not seem to support such a small value of $k_*$ we prefer to marginalise over this parameter.

The contours of $D_V(z_{\rm eff}) - \Omega_mh^2$ are shown in Figure~\ref{fig:chi2}, together with two degeneracy predictions~\citep{Eisenstein:2005su}. The solid line is that of constant $\Omega_mh^2D_V(z_{\rm eff})$, which gives the direction of degeneracy for a pure CDM model, where only the shape of the correlation function contributes to the fit, without a BAO peak. The dashed line corresponds to a constant $r_s(z_d)/D_V(z_{\rm eff})$, which is the degeneracy if only the position of the acoustic scale contributes to the fit. The dashed contours exclude the first data point, fitting from $20-190 h^{-1}$\;Mpc only, with the best fitting values $\alpha = 1.049\pm 0.071$ (corresponding to $D_V(z_{\rm eff}) = 462\pm31\;$Mpc), $\Omega_mh^2 = 0.129\pm0.025$ and $b=1.72\pm0.17$. The contours of this fit are tilted towards the dashed line, which means that the fit is now driven by the BAO peak, while the general fit (solid contours) seems to have some contribution from the shape of the correlation function. Excluding the first data point increases the error on the distance constraint only slightly from $5.9\%$ to $6.8\%$. The value of $\Omega_mh^2$ tends to be smaller, but agrees within $1\sigma$ with the former value.

Going back to the complete fit from $10-190h^{-1}\;$Mpc, we can include an external prior on $\Omega_mh^2$ from WMAP-7, which carries an error of only $4\%$ (compared to the $\approx 15\%$ we obtain by fitting our data). Marginalising over $\Omega_mh^2$ now gives $D_V(z_{\rm eff}) = 459\pm18\;$Mpc, which reduces the error from $5.9\%$ to $3.9\%$. The uncertainty in $\Omega_mh^2$ from WMAP-7 contributes only about $5\%$ of the error in $D_V$ (assuming no error in the WMAP-7 value of $\Omega_mh^2$ results in $D_V(z_{\rm eff}) = 459\pm17\;$Mpc).

In Figure~\ref{fig:plot5_sn} we plot the ratio $D_V(z)/D^{low-z}_V(z)$ as a function of redshift, where $D^{low-z}_V(z) = cz/H_0$. At sufficiently low redshift the approximation $H(z) \approx H_0$ is valid and the measurement is independent of any cosmological parameter except the Hubble constant. This figure also contains the results from~\citet{Percival:2009xn}.

Rather than including the WMAP-7 prior on $\Omega_mh^2$ to break the degeneracy between $\Omega_mh^2$ and the distance constraint, we can fit the ratio $r_s(z_d)/D_V(z_{\rm eff})$, where $r_s(z_d)$ is the sound horizon at the baryon drag epoch $z_d$. In principle, this is rotating Figure~\ref{fig:chi2} so that the dashed black line is parallel to the x-axis and hence breaks the degeneracy if the fit is driven by the BAO peak; it will be less efficient if the fit is driven by the shape of the correlation function. During the fit we calculate $r_s(z_d)$ using the fitting formula of~\citet{Eisenstein:1997ik}.

The best fit results in $r_s(z_d)/D_V(z_{\rm eff}) = 0.336\pm0.015$, which has an error of $4.5\%$,  smaller than the $5.9\%$ found for $D_V$ but larger than the error in $D_V$ when adding the WMAP-7 prior on $\Omega_mh^2$. This is caused by the small disagreement in the $D_V-\Omega_mh^2$ degeneracy and the line of constant sound horizon in Figure~\ref{fig:chi2}. The $\chi^2$ is $15.7$, similar to the previous fit with the same number of degrees of freedom.

\subsection{Extracting $A(z_{\rm eff})$ and $R(z_{\rm eff})$}
\label{sec:AandR}

We can also fit for the ratio of the distance between the effective redshift, $z_{\rm eff}$, and the redshift of decoupling \citep[$z_* = 1091$; ][]{Eisenstein:2005su}; 
\begin{equation}
R(z_{\rm eff}) = \frac{D_V(z_{\rm eff})}{(1+z_*)D_A(z_*)} , 
\end{equation}
with $(1+z_*)D_A(z_*)$ being the CMB angular comoving distance. Beside the fact that the Hubble constant $H_0$ cancels out in the determination of $R$, this ratio is also more robust against effects caused by  possible extra relativistic species~\citep{Eisenstein:2004an}. We calculate $D_A(z_*)$ for each $\Omega_mh^2$ during the fit and then marginalise over $\Omega_mh^2$. The best fit results in $R = 0.0324\pm 0.0015$, with $\chi^2 = 15.7$ and the same $14$ degrees of freedom.

Focusing on the path from $z=0$ to $z_{\rm eff}=0.106$, our dataset can give interesting constraints on $\Omega_m$. We derive the parameter~\citep{Eisenstein:2005su}
\begin{equation}
A(z_{\rm eff}) = 100D_V(z_{\rm eff})\frac{\sqrt{\Omega_mh^2}}{cz_{\rm eff}},
\label{eq:A}
\end{equation}
which has no dependence on the Hubble constant since $D_V \propto h^{-1}$. We obtain $A(z_{\rm eff}) = 0.526\pm0.028$ with $\chi^2/\rm d.o.f. = 15.7/14$. The value of $A$ would be identical to $\sqrt{\Omega_m}$ if measured at redshift $z=0$. At redshift $z_{\rm eff} = 0.106$ we obtain a deviation from this approximation of $6\%$ for our fiducial model, which is small but systematic. We can express $A$, including the curvature term $\Omega_k$ and the dark energy equation of state parameter $w$,  as
\begin{equation}
A(z) = \frac{\sqrt{\Omega_m}}{E(z)^{1/3}}
\begin{cases}
\left[\frac{\sinh\left(\sqrt{\Omega_k}\chi(z)\right)}{\sqrt{\Omega_k}z}\right]^{2/3} & \Omega_k > 0\cr
\left[\frac{\chi(z)}{z}\right]^{2/3} & \Omega_k = 0\cr
\left[\frac{\sin\left(\sqrt{|\Omega_k|}\chi(z)\right)}{\sqrt{|\Omega_k|}z}\right]^{2/3} & \Omega_k < 0
 \end{cases}
\end{equation}
with 
\begin{equation}
\chi(z) = D_C(z)\frac{H_0}{c} = \int^{z}_0\frac{dz'}{E(z')}
\end{equation}
and
\begin{align}
E(z) = \big[&\Omega_m(1+z)^3 + \Omega_k(1+z)^2\cr
             &+ \Omega_{\Lambda}(1+z)^{3(1+w)})\big]^{1/2} .
\end{align}
Using this equation we now linearise our result for $\Omega_m$ in $\Omega_k$ and $w$ and get
\begin{equation}
\Omega_m = 0.287 + 0.039(1+w) + 0.039\Omega_k \pm 0.027.
\end{equation}
For comparison, ~\citet{Eisenstein:2005su} found 
\begin{equation}
\Omega_m = 0.273 + 0.123(1+w) + 0.137\Omega_k \pm 0.025
\end{equation}
based on the SDSS LRG DR3 sample. This result shows the reduced sensitivity of the 6dFGS measurement to $w$ and $\Omega_k$.

\section{Cosmological implications}
\label{sec:results}

In this section we compare our results to other studies and discuss the implications for constraints on cosmological parameters. We first note that we do not see any excess correlation on large scales as found in the SDSS-LRG sample. Our correlation function is in agreement with a crossover to negative scales at $140h^{-1}$\;Mpc, as predicted from $\Lambda$CDM. 

\begin{table*}
\begin{center}
\caption{$w$CDM constraints from different datasets. Comparing the two columns shows the influence of the 6dFGS data point. The 6dFGS data point reduces the error on $w$ by $24\%$ compared to WMAP-7+LRG which contains only the BAO data points of~\citet{Percival:2009xn}. We assume flat priors of $0.11 < \Omega_mh^2 < 0.16$ and marginalise over $\Omega_mh^2$. The asterisks denote the free parameters in each fit.}
	\begin{tabular}{llll}
     		\hline
		parameter & WMAP-7+LRG & WMAP-7+LRG+6dFGS\\
		\hline
		$H_0$ & $69.9\pm3.8$(*) & $68.7\pm2.8$(*)\\
		$\Omega_m$ & $0.283\pm0.033$ & $0.293\pm0.027$\\
		$\Omega_{\Lambda}$ & $0.717\pm0.033$ & $0.707\pm0.027$\\
		$w$ & $\llap{-}1.01\pm0.17$(*) & $\llap{-}0.97\pm0.13$(*)\\
		\hline
	  \end{tabular}
	  \label{tab:comp}
\end{center}
\end{table*}

\begin{table*}
\begin{center}
\caption{Parameter constraints from WMAP7+BAO for (i) a flat $\Lambda$CDM model, (ii) an open $\Lambda$CDM (o$\Lambda$CDM), (iii) a flat model with $w = \rm const.$ ($w$CDM),  and (iv) an open model with $w = \rm constant$ (o$w$CDM). We assume flat priors of $0.11 < \Omega_mh^2 < 0.16$ and marginalise over $\Omega_mh^2$. The asterisks denote the free parameters in each fit.}
	\begin{tabular}{lllll}
     		\hline
		 parameter & $\Lambda$CDM & o$\Lambda$CDM & $w$CDM & o$w$CDM\\
		\hline
		$H_0$ & $69.2\pm1.1$(*) & $68.3\pm1.7$(*) & $68.7\pm2.8$(*) & $70.4\pm4.3$(*)\\
		$\Omega_m$ & $0.288\pm0.011$ & $0.290\pm0.019$ & $0.293\pm0.027$ & $0.274\pm0.035$\\
		$\Omega_{k}$ & ($0$) & $\llap{-}0.0036\pm0.0060$(*) & ($0$) & $\llap{-}0.013\pm0.010$(*)\\
		$\Omega_{\Lambda}$ & 0.$712\pm0.011$ & $0.714\pm0.020$ & $0.707\pm0.027$ & $0.726\pm0.036$\\
		$w$ & ($\;\llap{-}1$) & ($\;\llap{-}1$) & $\llap{-}0.97\pm0.13$(*) & $\llap{-}1.24\pm0.39$(*)\\
		\hline
	  \end{tabular}
	  \label{tab:models}
\end{center}
\end{table*}

\subsection{Constraining the Hubble constant, $H_0$}

\begin{figure}
\begin{center}
\epsfig{file=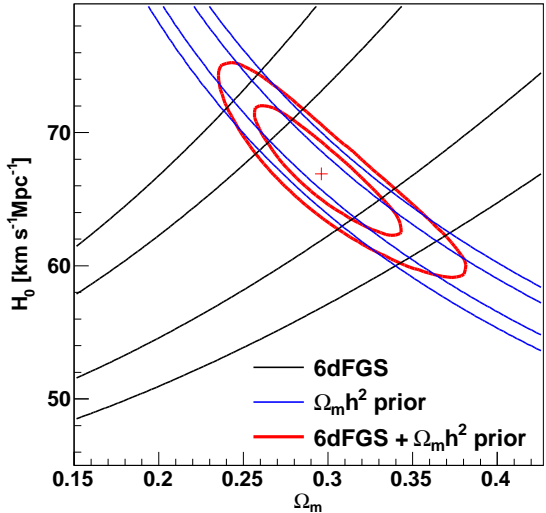,width=8cm}
\caption{The blue contours show the WMAP-7 $\Omega_mh^2$ prior~{\protect \citep{Komatsu:2010fb}}. The black contour shows constraints from 6dFGS derived by fitting to the measurement of $r_s(z_d)/D_V(z_{\rm eff})$. The solid red contours show the combined constraints resulting in $H_0 = 67\pm3.2\;$km\;s$^{-1}$Mpc$^{-1}$ and $\Omega_m = 0.296\pm0.028$. Combining the clustering measurement with $\Omega_mh^2$ from the CMB corresponds to the calibration of the standard ruler.}
\label{fig:chi2_h}
\end{center}
\end{figure}

\begin{figure}
\begin{center}
\epsfig{file=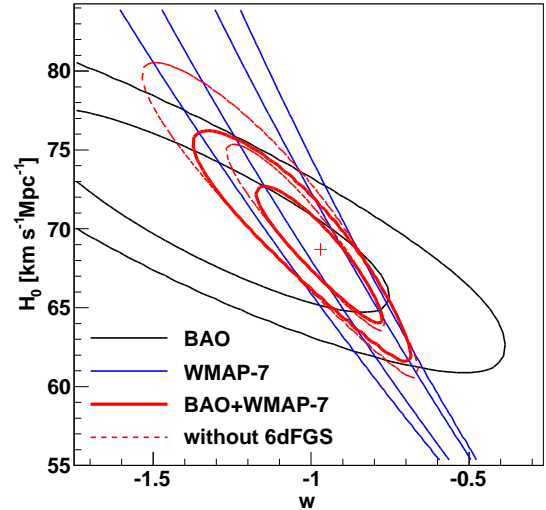,width=8cm}
\caption{The blue contours shows the WMAP-7 degeneracy in $H_0$ and $w$~{\protect \citep{Komatsu:2010fb}},
highlighting the need for a second dataset to break the degeneracy. The black contours show constraints from BAO data incorporating the $r_s(z_d)/D_V(z_{\rm eff})$ measurements of~{\protect \cite{Percival:2009xn}} and 6dFGS. The solid red contours show the combined constraints resulting in $w = -0.97\pm0.13$. Excluding the 6dFGS data point widens the constraints to the dashed red line with $w = -1.01\pm0.17$.}
\label{fig:chi2_w}
\end{center}
\end{figure}

We now use the 6dFGS data to derive an estimate of the Hubble constant. We use the 6dFGS measurement of $r_s(z_d)/D_V(0.106) = 0.336\pm0.015$ and fit directly for the Hubble constant and $\Omega_m$. We combine our measurement with a prior on $\Omega_mh^2$ coming from the WMAP-7 Markov chain results~\citep{Komatsu:2010fb}. Combining the clustering measurement with $\Omega_mh^2$ from the CMB corresponds to the calibration of the standard ruler.

We obtain values of $H_0 = 67\pm3.2\;$km s$^{-1}$Mpc$^{-1}$ (which has an uncertainty of only $4.8\%$) and $\Omega_m = 0.296\pm0.028$. Table~\ref{tab:para} and Figure~\ref{fig:chi2_h} summarise the results. The value of $\Omega_m$ agrees with the value we derived earlier (Section~\ref{sec:AandR}). 

To combine our measurement with the latest CMB data we use the WMAP-7 distance priors, namely the acoustic scale
\begin{equation}
\ell_A = (1+z_*)\frac{\pi D_A(z_*)}{r_s(z_*)},
\end{equation}
the shift parameter
\begin{equation}
R = 100\frac{\sqrt{\Omega_mh^2}}{c}(1+z_*)D_A(z_*)
\end{equation}
and the redshift of decoupling $z_*$ (Tables 9 and 10 in~\citealt{Komatsu:2010fb}). This combined analysis reduces the error further and yields $H_0 = 68.7\pm1.5\;$km s$^{-1}$Mpc$^{-1}$ ($2.2\%$) and $\Omega_m = 0.29\pm0.022$ ($7.6\%$).

\citet{Percival:2009xn} determine a value of $H_0 = 68.6\pm2.2$\;km s$^{-1}$Mpc$^{-1}$ using SDSS-DR7, SDSS-LRG and 2dFGRS, while \cite{Reid:2009xm} found $H_0 = 69.4\pm1.6$\;km\;s$^{-1}$Mpc$^{-1}$ using the SDSS-LRG sample and WMAP-5. In contrast to these results, 6dFGS is less affected by parameters like $\Omega_k$ and $w$ because of its lower redshift. In any case, our result of the Hubble constant agrees very well with earlier BAO analyses. Furthermore our result agrees with the latest CMB measurement of $H_0 = 70.3\pm2.5$\;km\;s$^{-1}$Mpc$^{-1}$~\citep{Komatsu:2010fb}.

The SH0ES program~\citep{Riess:2011yx} determined the Hubble constant using the distance ladder method. They used about $600$ near-IR observations of Cepheids in eight galaxies to improve the calibration of $240$ low redshift ($z < 0.1$) SN Ia, and  calibrated the Cepheid distances using the geometric distance to the maser galaxy NGC 4258. They found $H_0 = 73.8\pm2.4$\;km\;s$^{-1}$Mpc$^{-1}$, a value consistent with the initial results of the Hubble Key project \citet[$H_0 = 72\pm8$\;km\;s$^{-1}$Mpc$^{-1}$; ][]{Freedman:2000cf} but $1.7\sigma$ higher than our value (and $1.8\sigma$ higher when we combine our dataset with WMAP-7). While this could point toward unaccounted or under-estimated systematic errors in either one of the methods, the likelihood of such a deviation by chance is about $10\%$ and hence is not enough to represent a significant discrepancy. Possible systematic errors affecting the BAO measurements are the modelling of non-linearities, bias and redshift-space distortions, although these systematics are not expected to be significant at the large scales relevant to our analysis.

To summarise the finding of this section we can state that our measurement of the Hubble constant is competitive with the latest result of the  distance ladder method. The different techniques employed to derive these results have very different potential systematic errors. Furthermore we found that BAO studies provide the most accurate measurement of $H_0$ that exists, when combined with the CMB distance priors.

\subsection{Constraining dark energy}

One key problem driving current cosmology is the determination of the dark energy equation of state parameter, $w$. When adding additional parameters like $w$ to $\Lambda$CDM we find large degeneracies in the WMAP-7-only data. One example is shown in Figure~\ref{fig:chi2_w}. WMAP-7 alone can not constrain $H_0$ or $w$ within sensible physical boundaries (e.g. $w < -1/3$). As we are sensitive to $H_0$, we can break the degeneracy between $w$ and $H_0$ inherent in the CMB-only data. Our assumption of a fiducial cosmology with $w=-1$ does not introduce a bias, since our data is not sensitive to this parameter and any deviation from this assumption is modelled within the shift parameter $\alpha$.

We again use the WMAP-7 distance priors introduced in the last section. In addition to our value of $r_s(z_d)/D_V(0.106) = 0.336\pm0.015$ we use the results of~\citet{Percival:2009xn}, who found $r_s(z_d)/D_V(0.2) = 0.1905\pm0.0061$ and $r_s(z_d)/D_V(0.35) = 0.1097\pm0.0036$. To account for the correlation between the two latter data points we employ the covariance matrix reported in their paper. Our fit has $3$ free parameters, $\Omega_mh^2$, $H_0$ and $w$.

The best fit gives $w = -0.97\pm0.13$, $H_0 = 68.7\pm 2.8$\;km\;s$^{-1}$Mpc$^{-1}$ and $\Omega_mh^2 = 0.1380\pm 0.0055$, with a $\chi^2/\rm d.o.f. = 1.3/3$. Table~\ref{tab:comp} and Figure~\ref{fig:chi2_w} summarise the results.
To illustrate the importance of the 6dFGS result to the overall fit we also show how the results change if 6dFGS is omitted. The 6dFGS data improve the constraint on $w$ by $24\%$.

Finally we show the best fitting cosmological parameters for different cosmological models using WMAP-7 and BAO results in Table~\ref{tab:models}.

\section{Significance of the BAO detection}
\label{sec:sig}

\begin{figure}
\begin{center}
\epsfig{file=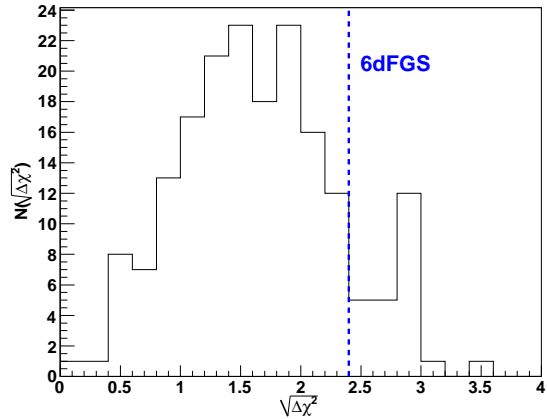,width=8cm}
\caption{The number of log-normal realisations found with a certain $\sqrt{\Delta\chi^2}$, where the $\Delta\chi^2$ is obtained by comparing a fit using a $\Lambda$CDM correlation function model with a no-baryon model. The blue line indicates the 6dFGS result.}
\label{fig:sigma}
\end{center}
\end{figure}

\begin{figure}
\begin{center}
\epsfig{file=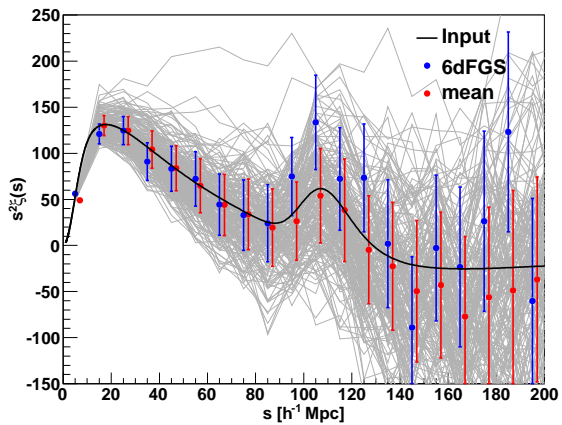,width=8cm}
\caption{The different log-normal realisations used to calculate the covariance matrix (shown in grey). The red points indicate the mean values, while the blue points show actual 6dFGS data (the data point at $5h^{-1}\;$Mpc is not included in the fit). The red data points are shifted by $2h^{-1}$\;Mpc to the right for clarity.}
\label{fig:log_6df}
\end{center}
\end{figure}

\begin{figure}
\begin{center}
\epsfig{file=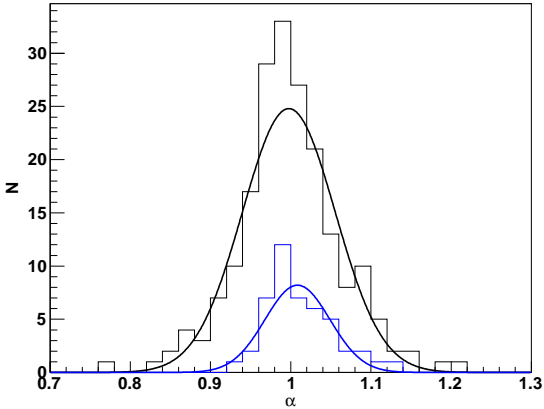,width=8cm}
\caption{This plot shows the distribution of the parameter $\alpha$ derived from the $200$ log-normal realisations (black). The distribution is well fit by a Gaussian with a mean of $\mu = 0.998\pm0.004$ and a width of $\sigma = 0.057\pm0.005$. In blue we show the same distribution selecting only the log-normal realisations with a strong BAO peak ($> 2\sigma$). The Gaussian distribution in this case gives a mean of $1.007\pm0.007$ and $\sigma=0.041\pm0.008$.}
\label{fig:alpha_distri}
\end{center}
\end{figure}

To test the significance of our detection of the BAO signature we follow~\citet{Eisenstein:2005su} and perform a fit with a fixed $\Omega_b = 0$, which corresponds to a pure CDM model without a BAO signature. The best fit has $\chi^2 = 21.4$ with $14$ degrees of freedom and is shown as the red dashed line in Figure~\ref{fig:bao}. The parameter values of this fit depend on the parameter priors, which we set to $0.7 < \alpha < 1.3$ and $0.1 < \Omega_mh^2 < 0.2$. Values of $\alpha$ much further away from $1$ are problematic since eq.~\ref{eq:alpha} is only valid for $\alpha$ close to $1$. Comparing the best pure CDM model with our previous fit, we estimate that the BAO signal is detected with a significance of $2.4\sigma$ (corresponding to $\Delta\chi^2 = 5.6$). As a more qualitative argument for the detection of the BAO signal we would like to refer again to Figure~\ref{fig:chi2} where the direction of the degeneracy clearly indicates the sensitivity to the BAO peak.
 
We can also use the log-normal realisations to determine how likely it is to find a BAO detection in a survey like 6dFGS. To do this, we produced $200$ log-normal mock catalogues and calculated the correlation function for each of them. We can now fit our correlation function model to these realisations. Furthermore, we fit a no-baryon model to the correlation function and calculate $\Delta\chi^2$, the distribution of which is shown in Figure~\ref{fig:sigma}. We find that $26\%$ of all realisations have {\sl at least} a $2\sigma$ BAO detection, and that  $12\%$ have a detection $> 2.4\sigma$. The log-normal realisations show a mean significance of the BAO detection of $1.7\pm0.7\sigma$, where the error describes the variance around the mean.

Figure~\ref{fig:log_6df} shows the 6dFGS data points together with all $200$ log-normal realisations (grey). The red data points indicate the mean for each bin and the black line is the input model derived as explained in Section~\ref{sec:log}. This comparison shows that the 6dFGS data contain a BAO peak slightly larger than expected in $\Lambda$CDM.

The amplitude of the acoustic feature relative to the overall normalisation of the galaxy correlation function is quite sensitive to the baryon fraction, $f_b = \Omega_b/\Omega_m$~\citep{Matsubara:2004fr}. A higher BAO peak could hence point towards a larger baryon fraction in the local universe. However since the correlation function model seems to agree very well with the data (with a reduced $\chi^2$ of $1.12$) and is within the range spanned by our log-normal realisations, we can not claim any discrepancy with $\Lambda$CDM. Therefore, the most likely explanation for the excess correlation in the BAO peak is sample variance.

In Figure~\ref{fig:alpha_distri} we show the distribution of the parameter $\alpha$ obtained from the $200$ log-normal realisations. The distribution is well described by a Gaussian with $\chi^2/\text{d.o.f.} = 14.2/20$, where we employed Poisson errors for each bin. This confirms that $\alpha$ has Gaussian distributed errors in the approximation that the 6dFGS sample is well-described by log-normal realisations of an underlying $\Lambda$CDM power spectrum. This result increases our confidence that the application of Gaussian errors for the cosmological parameter fits is correct. The mean of the Gaussian distribution is at $0.998\pm0.004$ in agreement with unity, which shows, that we are able to recover the input model. The width of the distribution shows the mean expected error in $\alpha$ in a $\Lambda$CDM universe for a 6dFGS-like survey. We found $\sigma = 0.057\pm0.005$ which is in agreement with our error in $\alpha$ of $5.9\%$. Figure~\ref{fig:alpha_distri} also contains the distribution of $\alpha$, selecting only the log-normal realisations with a strong ($>2\sigma$) BAO peak (blue data). We included this selection to show, that a stronger BAO peak does not bias the estimate of $\alpha$ in any direction. The Gaussian fit gives $\chi^2/\text{d.o.f.} = 5/11$ with a mean of $1.007\pm0.007$. The distribution of $\alpha$ shows a smaller spread with $\sigma = 0.041\pm0.008$, about $2\sigma$ below our error on $\alpha$. This result shows, that a survey like 6dFGS is able to constrain $\alpha$ (and hence $D_V$ and $H_0$) to the precision we report in this paper.

\section{Future all sky surveys}
\label{sec:future}

\begin{figure}
\begin{center}
\epsfig{file=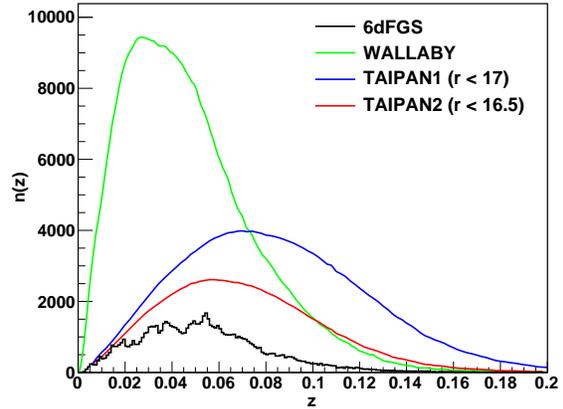,width=8cm}
\caption{Redshift distribution of 6dFGS, WALLABY and two different versions of the proposed TAIPAN survey.
See text for details.}
\label{fig:red_comp}
\end{center}
\end{figure}

\begin{figure}
\begin{center}
\epsfig{file=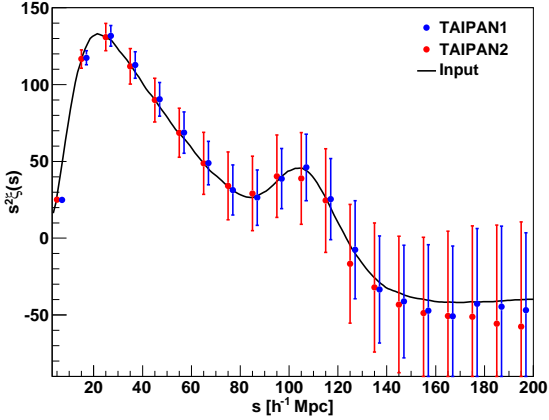,width=8cm}
\caption{Predictions for two versions of the proposed TAIPAN survey. Both predictions assume a $2\pi$\;steradian southern sky-coverage, excluding the Galactic plane (i.e. $|b| > 10^\circ$). TAIPAN1 contains $406\,000$ galaxies while TAIPAN2 contains $221\,000$, (see Figure~\ref{fig:red_comp}). The blue points are shifted by $2h^{-1}$\;Mpc to the right for clarity. The black line is the input model, which is a $\Lambda$CDM model with a bias of $1.6$, $\beta = 0.3$ and $k_* = 0.17h\;$Mpc$^{-1}$. For a large number of realisations, the difference between the input model and the mean (the data points) is only the convolution with the window function.}
\label{fig:log_taipan_both}
\end{center}
\end{figure}

\begin{figure}
\begin{center}
\epsfig{file=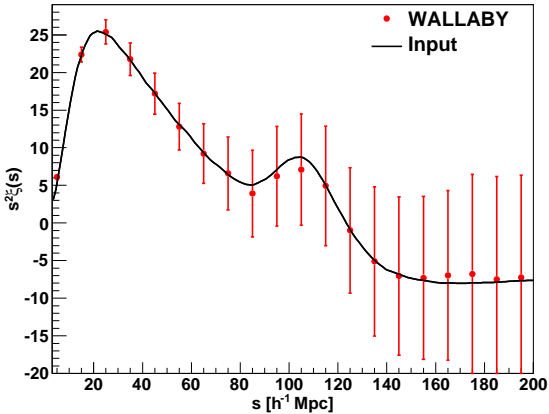,width=8cm}
\caption{Prediction for the WALLABY survey. We have assumed a $4\pi$\;steradian survey with $602\,000$ galaxies, $b = 0.7$, $\beta = 0.7$ and $k_* = 0.17h$\;Mpc$^{-1}$.}
\label{fig:log_wallaby}
\end{center}
\end{figure}

A major new wide-sky survey of the local Universe will be the Wide field ASKAP L-band Legacy All-sky Blind surveY (WALLABY)\footnote{http://www.atnf.csiro.au/research/WALLABY}. This is a blind H{\sc I} survey planned for the Australian SKA Pathfinder telescope (ASKAP), currently under construction at the Murchison Radio-astronomy Observatory (MRO) in Western Australia.

The survey will cover at least $75\%$ of the sky with the potential to cover $4\pi$ of sky if the Westerbork Radio Telescope delivers complementary northern coverage. Compared to 6dFGS, WALLABY will more than double the sky coverage including the Galactic plane. WALLABY will contain $\sim 500\,000$ to $600\,000$ galaxies with a mean redshift of around $0.04$, giving it around 4 times greater galaxy density compared to 6dFGS. In the calculations that follow, we assume for WALLABY a $4\pi$ survey without any exclusion around the Galactic plane. The effective volume in this case turns out to be $0.12h^{-3}$\;Gpc$^3$.

The TAIPAN survey\footnote{TAIPAN: Transforming Astronomical Imaging surveys through Polychromatic Analysis of Nebulae} proposed for the UK Schmidt Telescope at Siding Spring Observatory, will cover a comparable area of sky, and will extend 6dFGS in both depth and redshift ($z\simeq 0.08$).

The redshift distribution of both surveys is shown in Figure~\ref{fig:red_comp}, alongside 6dFGS. Since the TAIPAN survey is still in the early planning stage we consider two realisations: TAIPAN1 ($406\,000$ galaxies to a faint magnitude limit of $r=17$) and the shallower TAIPAN2 ($221\,000$ galaxies to $r=16.5$). We have adopted the same survey window as was used for 6dFGS, meaning that it covers the whole southern sky excluding a $10^{\circ}$ strip around the Galactic plane. The effective volumes of TAIPAN1 and TAIPAN2 are $0.23h^{-3}$\;Gpc$^3$ and $0.13h^{-3}$\;Gpc$^3$, respectively.

To predict the ability of  these surveys to measure the large scale correlation function we produced $100$ log-normal realisations for TAIPAN1 and WALLABY and $200$ log-normal realisations for TAIPAN2. Figures~\ref{fig:log_taipan_both} and \ref{fig:log_wallaby} show the results in each case. The data points are the mean of the different realisations, and the error bars are the diagonal of the covariance matrix. The black line represents the input model which is a $\Lambda$CDM prediction convolved with a Gaussian damping term using $k_* = 0.17h$\;Mpc$^{-1}$ (see eq.~\ref{eq:damp}). We used a bias parameter of $1.6$ for TAIPAN and following our fiducial model we get $\beta = 0.3$, resulting in $A = b^2(1 + 2\beta/3 + \beta^2/5) = 3.1$. For WALLABY we used a bias of $0.7$ ~\citep[based on the results found in the HIPASS survey; ][]{Basilakos:2007wq}. This results in $\beta = 0.7$ and $A = 0.76$. To calculate the correlation function we used $P_0 = 40\,000h^3\;$Mpc$^3$ for TAIPAN and $P_0 = 5\,000h^3\;$Mpc$^3$ for WALLABY.

The error bar for TAIPAN1 is smaller by roughly a factor of $1.7$ relative to 6dFGS, which is consistent with scaling by $\sqrt{V_{\rm eff}}$ and is comparable to the SDSS-LRG sample. We calculate the significance of the BAO detection for each log-normal realisation by performing fits to the correlation function using $\Lambda$CDM parameters and $\Omega_b = 0$, in exactly the same manner as the 6dFGS analysis described earlier. We find a $3.5\pm0.8\sigma$ significance for the BAO detection for TAIPAN1, $2.1\pm0.7\sigma$ for TAIPAN2 and $2.1\pm0.7\sigma$ for WALLABY, where the error again describes the variance around the mean. 

We then fit a correlation function model to the mean values of the log-normal realisations for each survey, using the covariance matrix derived from these log-normal realisations. We evaluated the correlation function of WALLABY, TAIPAN2 and TAIPAN1 at the effective redshifts of $0.1$, $0.12$ and $0.14$, respectively. With these in hand, we are able to derive distance constraints to respective precisions of $7\%$, $6\%$ and $3\%$. The predicted value for WALLABY is not significantly better than that from 6dFGS. This is due to the significance of the 6dFGS BAO peak in the data, allowing us to place tight constraints on the distance. As an alternative figure-of-merit, we derive the constraints on the Hubble constant. All surveys recover the input parameter of $H_0 = 70\;$km\;s$^{-1}$Mpc$^{-1}$, with absolute uncertainties of $3.7$, $3$ and $2.2\;$km\;s$^{-1}$Mpc$^{-1}$ for WALLABY, TAIPAN2 and TAIPAN1, respectively. Hence, TAIPAN1 is able to constrain the Hubble constant to $3\%$ precision. These constraints might improve when combined with Planck constraints on $\Omega_bh^2$ and $\Omega_mh^2$ which will be available when these surveys come along. 

Since there is significant overlap between the survey volume of 6dFGS, TAIPAN and WALLABY, it might be interesting to test whether the BAO analysis of the local universe can make use of a multiple tracer analysis, as suggested recently by~\citet{ArnalteMur:2011xp}. These authors claim that by employing {\sl two} different tracers of the matter density field --
one with high bias to trace the central over-densities, and one with low bias to trace the small density fluctuations -- one 
can improve the detection and measurement of the BAO signal. \citet{ArnalteMur:2011xp} test this approach using the SDSS-LRG sample (with a very large bias) and the SDSS-main sample (with a low bias). Although the volume is limited by the amount of sample overlap, they detect the BAO peak at $4.1\sigma$. Likewise, we expect that the contrasting high bias of 6dFGS and TAIPAN, when used  in conjunction with the low bias of WALLABY, would furnish a combined sample that would be ideal for such an analysis. 

Neither TAIPAN nor WALLABY are designed as BAO surveys, with their primary goals relating to galaxy formation and
the local universe. However, we have found that TAIPAN1 would be able to improve the measurement of the local Hubble constant by about $30\%$ compared to 6dFGS going to only slightly higher redshift. WALLABY could make some interesting contributions in the form of a multiple tracer analysis.

\section{Conclusion}
\label{sec:conclusion}

We have calculated the large-scale correlation function of the 6dF Galaxy Survey and detected a BAO peak with a significance of $2.4\sigma$. Although 6dFGS was never designed as a BAO survey, the peak is detectable because the survey
contains a large number of very bright, highly biased galaxies, within a sufficiently large effective volume of $0.08h^{-3}$\;Gpc$^3$. We draw the following conclusions from our work:

\begin{itemize}
\item The 6dFGS BAO detection confirms the finding by SDSS and 2dFGRS of a peak in the correlation function at around $105h^{-1}$\;Mpc, consistent with $\Lambda$CDM. This is important because 6dFGS is an independent sample, with a different target selection,  redshift distribution, and bias compared to previous studies. The 6dFGS BAO measurement is the lowest redshift BAO measurement ever made.
\item We do not see any excess correlation at large scales as seen in the SDSS-LRG sample. Our correlation function is consistent with a crossover to negative values at $140h^{-1}$\;Mpc, as expected from $\Lambda$CDM models.
\item We derive the distance to the effective redshift as $D_V(z_{\rm eff}) = 456\pm27\;$Mpc ($5.9\%$ precision). Alternatively, we can derive $r_s(z_d)/D_V(z_{\rm eff}) = 0.336\pm0.015$ ($4.5\%$ precision). All parameter constraints are summarised in Table~\ref{tab:para}.
\item Using a prior on $\Omega_mh^2$ from WMAP-7, we find $\Omega_m = 0.296\pm0.028$. Independent of WMAP-7, and taking into account curvature and the dark energy equation of state, we derive $\Omega_m = 0.287 + 0.039(1+w) + 0.039\Omega_k \pm 0.027$. This agrees very well with the first value, and shows the very small dependence on cosmology for parameter derivations from 6dFGS given its low redshift.
\item We are able to measure the Hubble constant, $H_0 = 67\pm3.2\;$km s$^{-1}$Mpc$^{-1}$, to $4.8\%$ precision, using only the standard ruler calibration by the CMB (in form of $\Omega_mh^2$ and $\Omega_bh^2$). Compared to previous BAO measurements, 6dFGS is almost completely independent of cosmological parameters (e.g. $\Omega_k$ and $w$), similar to Cepheid and low-$z$ supernovae methods. However, in contrast to these methods, the BAO derivation of the Hubble constant depends on very basic early universe physics and avoids possible systematic errors coming from the build up of a distance ladder.
\item By combining the 6dFGS BAO measurement with those of  WMAP-7 and previous redshift samples \citet[from SDSS-DR7, SDDS-LRG and 2dFGRS; ][]{Percival:2009xn}, we can further improve the constraints on the dark energy equation of state, $w$, by breaking the $H_0-w$ degeneracy in the CMB data. Doing this, we find $w=-0.97\pm0.13$, which is an improvement of $24\%$ compared to previous combinations of BAO and WMAP-7 data.
\item We have made detailed predictions for two next-generation low redshift surveys,  WALLABY and TAIPAN. Using our 6dFGS result, we predict that both surveys will detect the BAO signal, and that WALLABY may be the first radio galaxy survey to do so. Furthermore, we predict that TAIPAN has the potential to constrain the Hubble constant to a precision of $3\%$ improving the 6dFGS measurement by $30\%$.
\end{itemize}

\section*{Acknowledgments}

The authors thank Alex Merson for providing the random mock generator and Lado Samushia for helpful advice with the wide-angle formalism. We thank Martin Meyer and Alan Duffy for fruitful discussions and Greg Poole for providing the relation for the scale dependent bias. We also thank Tamara Davis, Eyal Kazin and John Peacock for comments on earlier versions of this paper.
F.B. is supported by the Australian Government through the International Postgraduate Research Scholarship (IPRS) and by scholarships from ICRAR and the AAO. Part of this work used the ivec$@$UWA supercomputer facility.
The 6dF Galaxy Survey was funded in part by an Australian Research Council Discovery--Projects Grant (DP-0208876), administered by the Australian National University.

\setlength{\bibhang}{2em}
\setlength{\labelwidth}{0pt}

\appendix

\section{Generating log-normal mock catalogues}
\label{ap:log}

Here we explain in detail the different steps used to derive a log-normal mock catalogue, as a useful guide for researchers in the field. We start with an input power spectrum, (which is determined as explained in Section~\ref{sec:log}) in units of $h^{-3}$Mpc$^{3}$. We set up a 3D grid with the dimensions $L_x\times L_y\times L_z = 1000\times 1000\times 1000h^{-1}\;$Mpc with $200^3$ sub-cells. We then distribute the quantity $P(\vec{k})/V$ over this grid, where $V$ is the volume of the grid and $\vec{k} = \sqrt{k_x^2 + k_y^2 + k_z^2}$ with $k_x = n_x2\pi/L_x$ and $n_x$ being an integer value specifying the $x$ coordinates of the grid cells.

Performing a complex-to-real Fourier transform (FT) of this grid will produce a 3D correlation function. Since the power spectrum has the property $P(-\vec{k}) = P(\vec{k})^*$ the result will be real.

The next step is to replace the correlation function $\xi(r)$ at each point in the 3D grid by $\ln[1 + \xi(r)]$, where $\ln$ is the natural logarithm. This step prepares the input model for the inverse step, which we later use to produce the log-normal density field. 

Using a real-to-complex FT we can revert to $k$-space where we now have a modified power spectrum, $P_{\ln}(\vec{k})$. At this point we divide by the number of sub-cells $N_c$. The precise normalisation  depends on the definition of the discrete Fourier transform. We use the FFTW library~\citep{Frigo:2005}, where the discrete FT is defined as
\begin{equation}
Y_i = \sum^{N_c-1}_{j=0} X_j \exp\left[\pm 2\pi ij\sqrt{-1}/N_c\right].
\end{equation}
The modified power spectrum $P_{\ln}(\vec{k})$ is not guarantied to be neither positive defined nor a real function, which contradicts the definition of a power spectrum. \citet{Weinberg:1991qe} suggested to construct a well defined power spectrum from $P_{\ln}(\vec{k})$ by
\begin{equation}
P'_{\ln}(\vec{k}) = \text{max}\left[0,\text{Re}[P_{\ln}(\vec{k})]\right].
\end{equation}
We now generate a real and an imaginary Fourier amplitude $\delta(\vec{k})$ for each point on the grid by randomly sampling from a Gaussian distribution with r.m.s. $\sqrt{P'_{\ln}(\vec{k})/2}$. However, to ensure that the final over-density field is real, we have to manipulate the grid, so that all sub-cells follow the condition $\delta(-\vec{k}) = \delta(\vec{k})^*$.

Performing another FT results in an over-density field $\delta(\vec{x})$ from which we calculate the variance $\sigma_G^2$. The mean of $\delta(\vec{x})$ should be zero. The log-normal density field is then given by
\begin{equation}
\mu_L(\vec{x}) = \exp\left[\delta(\vec{x}) - \sigma_G^2/2\right],
\end{equation}
which is now a quantity defined on $[0,\infty[$ only, while $\delta(\vec{x})$ is defined on $]-\infty,\infty[$.

Since we want to calculate a mock catalogue for a particular survey we have to incorporate the survey selection function. If $W(\vec{x})$ is the selection function with the normalisation $\sum W(\vec{x}) = 1$, we calculate the mean number of galaxies in each grid cell as 
\begin{equation}
n_g(\vec{x}) = N\;W(\vec{x})\;\mu_L(\vec{x}),
\end{equation}
where $N$ is the total number of galaxies in our sample. The galaxy catalogue itself is than generated by Poisson sampling $n_g(\vec{x})$.

The galaxy position is not defined within the sub-cell, and we place the galaxy in a random position within the box. This means that the correlation function calculated from such a distribution is smooth at scales smaller than the sub-cell. It is therefore important to make sure that the grid cells are smaller than the size of the bins in the correlation function calculation. In the 6dFGS calculations presented in this paper the grid cells have a size of $5h^{-1}\;$Mpc, while the correlation function bins are $10h^{-1}\;$Mpc in size.

\section{Comparison of log-normal and jack-knife error estimates}
\label{ap:jk_comp}

\begin{figure}
\begin{center}
\epsfig{file=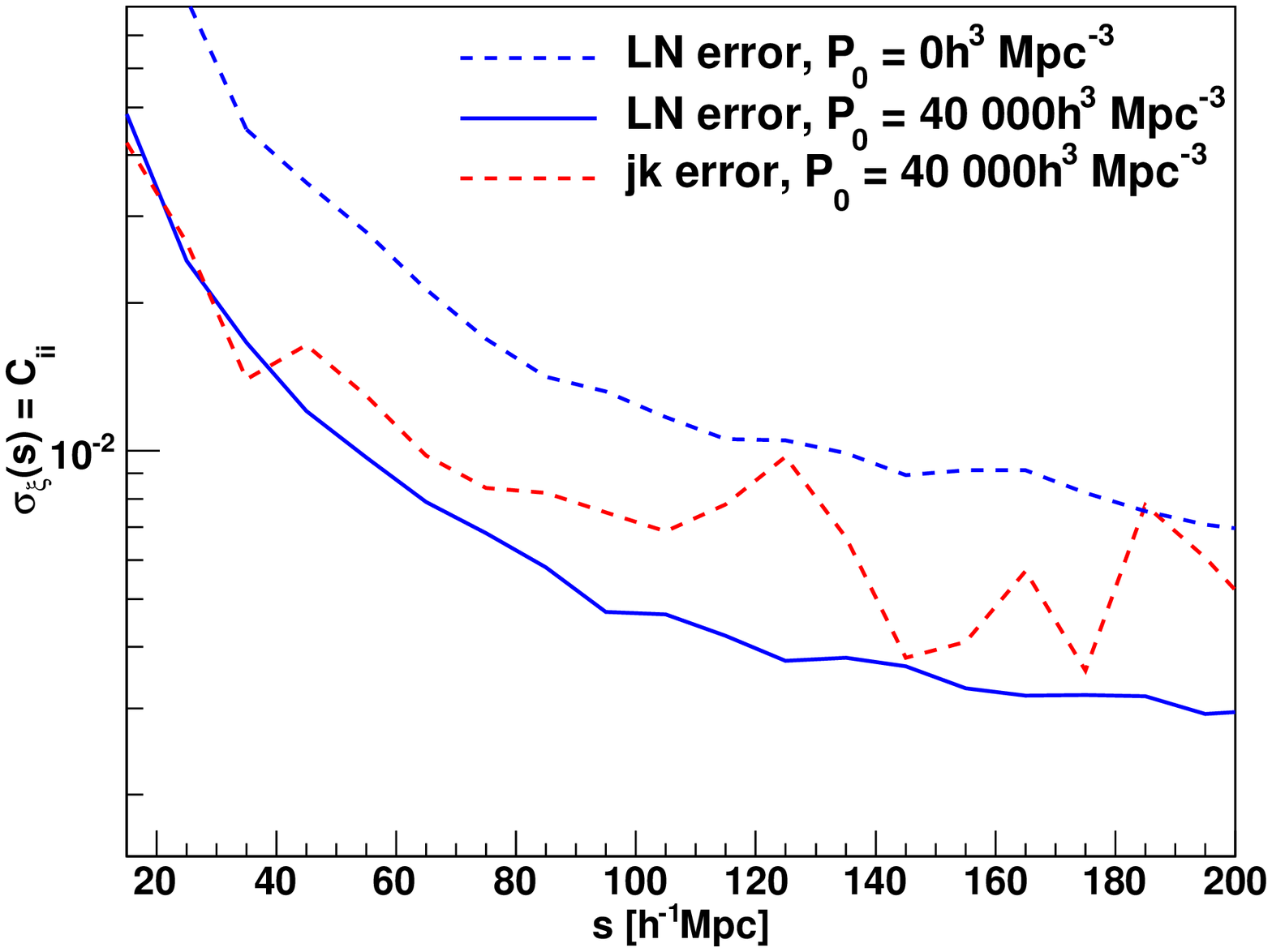, width=8cm}
\caption{Correlation function error for different values of $P_0$. The weighting with $P_0 = 40\,000h^{3}\;$Mpc$^{-3}$ reduces the error at the BAO scale by almost a factor of four compared to the case without weighting.  The red dashed line indicates the jack-knife error.}
\label{fig:sigP0}
\end{center}
\end{figure}

\begin{figure}
\begin{center}
\epsfig{file=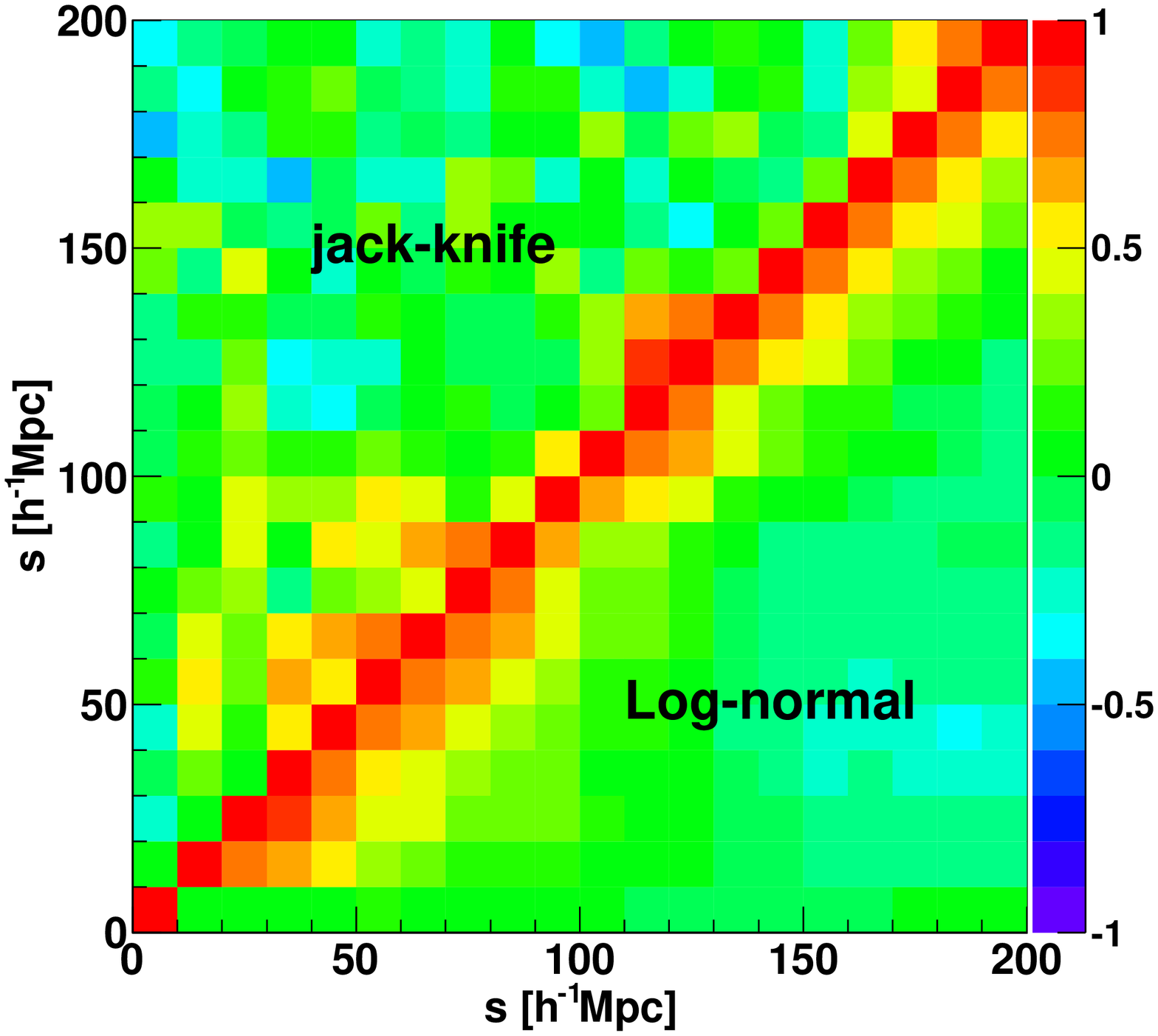,width=8cm}
\caption{Correlation matrix of the jack-knife errors (upper left triangle) and log-normal errors (lower right triangle).}
\label{fig:matrix2}
\end{center}
\end{figure}

We have also estimated jack-knife errors for the correlation function, by way of comparison. We divided the survey into $18$ regions and calculated the correlation function by excluding one region at a time. We found that the size of the error-bars around the BAO peak varies by around $20\%$ in some bins, when we increase the number of jack-knife regions from $18$ to $32$. Furthermore the covariance matrix derived from jack-knife resampling is very noisy and hard to invert.

We show the jack-knife errors in Figure~\ref{fig:sigP0}. The jack-knife error shows more noise and is larger in most bins compared to the log-normal error. The error shown in Figure~\ref{fig:sigP0} is only the diagonal term of the covariance matrix and does not include any correlation between bins.

The full error matrix is shown in Figure~\ref{fig:matrix2}, where we plot the correlation matrix of the jack-knife error estimate compared to the log-normal error. The jack-knife correlation matrix looks much more noisy and seems to have less correlation in neighbouring bins.

The number of jack-knife regions can not be chosen arbitrarily. Each jack-knife region must be at least as big as the maximum scale under investigation. Since we want to test scales up to almost $200h^{-1}\;$Mpc our jack-knife regions must be very large. On the other hand we need at least as many jack-knife regions as we have bins in our correlation function, otherwise the covariance matrix is singular. These requirements can contradict each other, especially if large scales are analysed. Furthermore the small number of jack-knife regions is the main source of noise (for a more detailed study of jack-knife errors see e.g.~\citealt{Norberg:2008tg}).

Given these limitations in the jack-knife error approach, correlation function studies on large scales usually employ simulations or log-normal realisations to derive the covariance matrix. We decided to use the log-normal error in our analysis. We showed that the jack-knife errors tend to be larger than the log-normal error at larger scales and carry less correlation. These differences might be connected to the much higher noise level in the jack-knife errors, which is clearly visible in all our data. It could be, however, that our jack-knife regions are too small to deliver reliable errors on large scales. We use the minimum number of jack-knife regions to make the covariance matrix non-singular (the correlation function is measured in $18$ bins). The mean distance of the jack-knife regions to each other is about $200h^{-1}\;$Mpc at the mean redshift of the survey, but smaller at low redshift.

\section{Wide-angle formalism}
\label{ap:wide}

The general redshift space correlation function (ignoring the plane parallel approximation) depends on $\phi$, $\theta$ and $s$. Here, $s$ is the separation between the galaxy pair, $\theta$ is the half opening angle, and $\phi$ is the angle of $s$ to the line of sight (see Figure~$1$ in~\citealt{Raccanelli:2010hk}). For the following calculations it must be considered that in this parametrisation, $\phi$ and $\theta$ are not independent.

The total correlation function model, including $O(\theta^2)$ correction terms, is then given by~\cite{Papai:2008bd},
\begin{equation}
\begin{split}
\xi_s(\phi,\theta,s) &= a_{00} + 2a_{02}\cos(2\phi) + a_{22}\cos(2\phi) + b_{22}\sin^2(2\phi)\\
& +\Big[ - 4a_{02}\cos(2\phi) - 4a_{22} - 4b_{22} - 4a_{10}\cot^2(\phi)\\
& + 4a_{11}\cot^2(\phi) - 4a_{12}\cot^2(\phi)\cos(2\phi) + 4b_{11}\\
& - 8b_{12}\cos^2(\phi)\Big]\theta^2 + O(\theta^4)
\end{split} 
\label{eq:wide1}
\end{equation}
This equation reduces to the plane parallel approximation if $\theta = 0$. The factors $a_{xy}$ and $b_{xy}$ in this equation are given by
\begin{equation}
\begin{split}
a_{00} &= \left[1 + \frac{2\beta}{3} + \frac{2\beta^2}{15}\right]\xi_0^2(r)\\
&- \left[\frac{\beta}{3} + \frac{2\beta^2}{21}\right]\xi^2_2(r) + \frac{3\beta^2}{140}\xi^2_4(r)\\
a_{02} &= -\left[\frac{\beta}{2} + \frac{3\beta^2}{14}\right]\xi^2_2(r) + \frac{\beta^2}{28}\xi^2_4(r)\\
a_{22} &= \frac{\beta^2}{15}\xi^2_0(r) - \frac{\beta^2}{21}\xi^2_2(r) + \frac{19\beta^2}{140}\xi^2_4(r)\\
b_{22} &= \frac{\beta^2}{15}\xi_0^2(r) - \frac{\beta^2}{21}\xi^2_2(r) - \frac{4\beta^2}{35}\xi^2_4(r)\\
a_{10} &= \left[2\beta + \frac{4\beta^2}{5}\right]\frac{1}{r}\xi^1_1(r) - \frac{\beta^2}{5r}\xi^1_3(r)\\
a_{11} &= \frac{4\beta^2}{3r^2}\left[\xi^0_0(r) - 2\xi^0_2(r)\right]\\
a_{21} &= \frac{\beta^2}{5r}\left[3\xi^1_3(r) - 2\xi^1_1(r)\right]\\
b_{11} &= \frac{4\beta^2}{3r^2}\left[\xi^0_0(r) + \xi^0_2(r)\right]\\
b_{12} &= \frac{2\beta^2}{5r}\left[\xi^1_1(r) + \xi^1_3(r)\right] ,
\end{split}
\end{equation}
where $\beta = \Omega_m(z)^{0.545}/b$, with $b$ being the linear bias. The correlation function moments are given by 
\begin{equation}
\xi^m_l(r) = \frac{1}{2\pi^2}\int^{\infty}_0 dk\;k^mP_{\rm lin}(k)j_l(rk)
\end{equation}
with $j_l(x)$ being the spherical Bessel function of order $l$.

\begin{figure}
\begin{center}
\epsfig{file=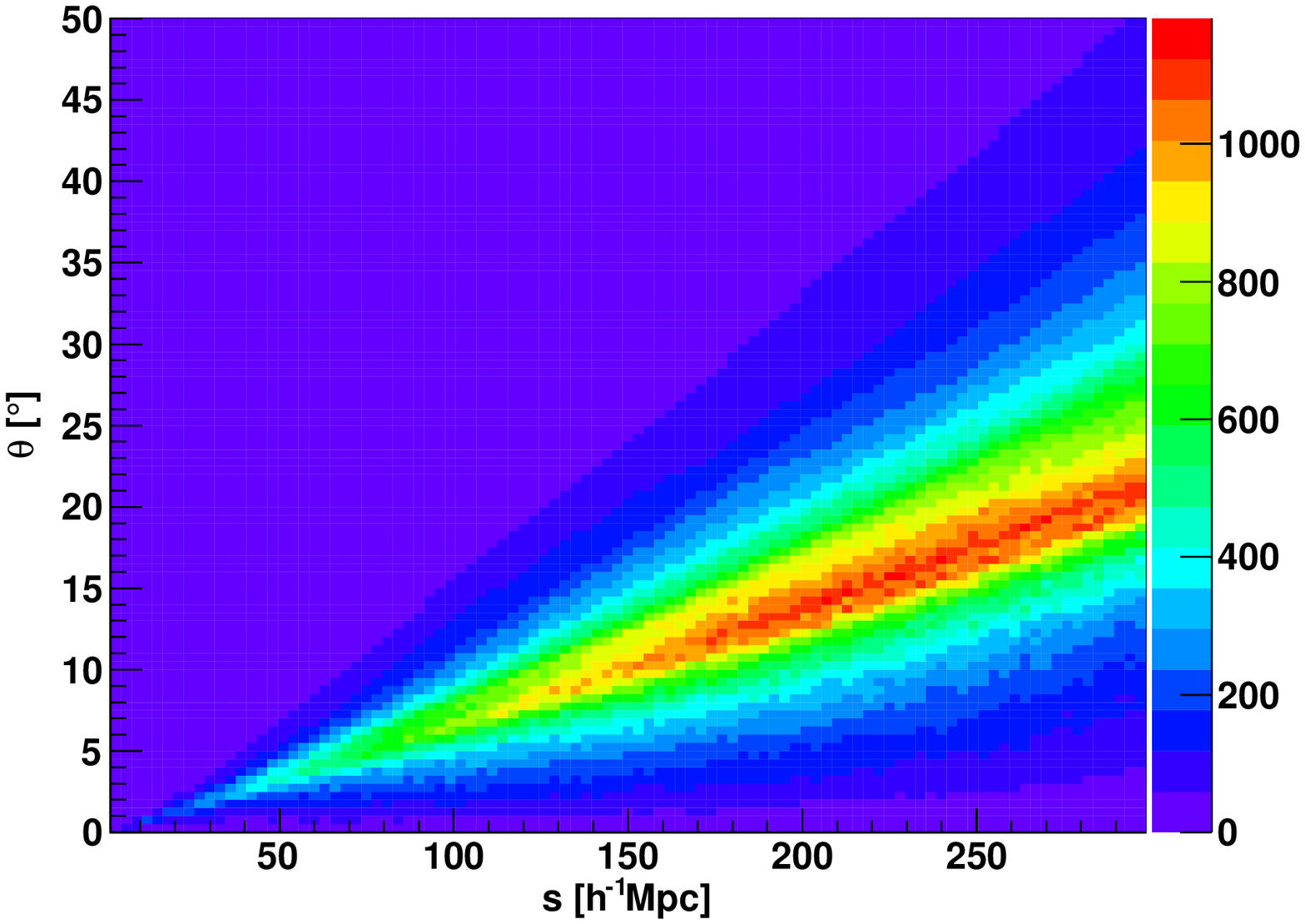, width=8cm}
\caption{The half opening angle $\theta$ as a function of separation $s$ of the 6dFGS weighted catalogue. The plane parallel approximation assumes $\theta = 0$. The mean half opening angle at the BAO scale is $\lesssim 10^{\circ}$. The colour bar gives the number of pairs in each bin.}
\label{fig:theta_s}
\end{center}
\end{figure}

\begin{figure}
\begin{center}
\epsfig{file=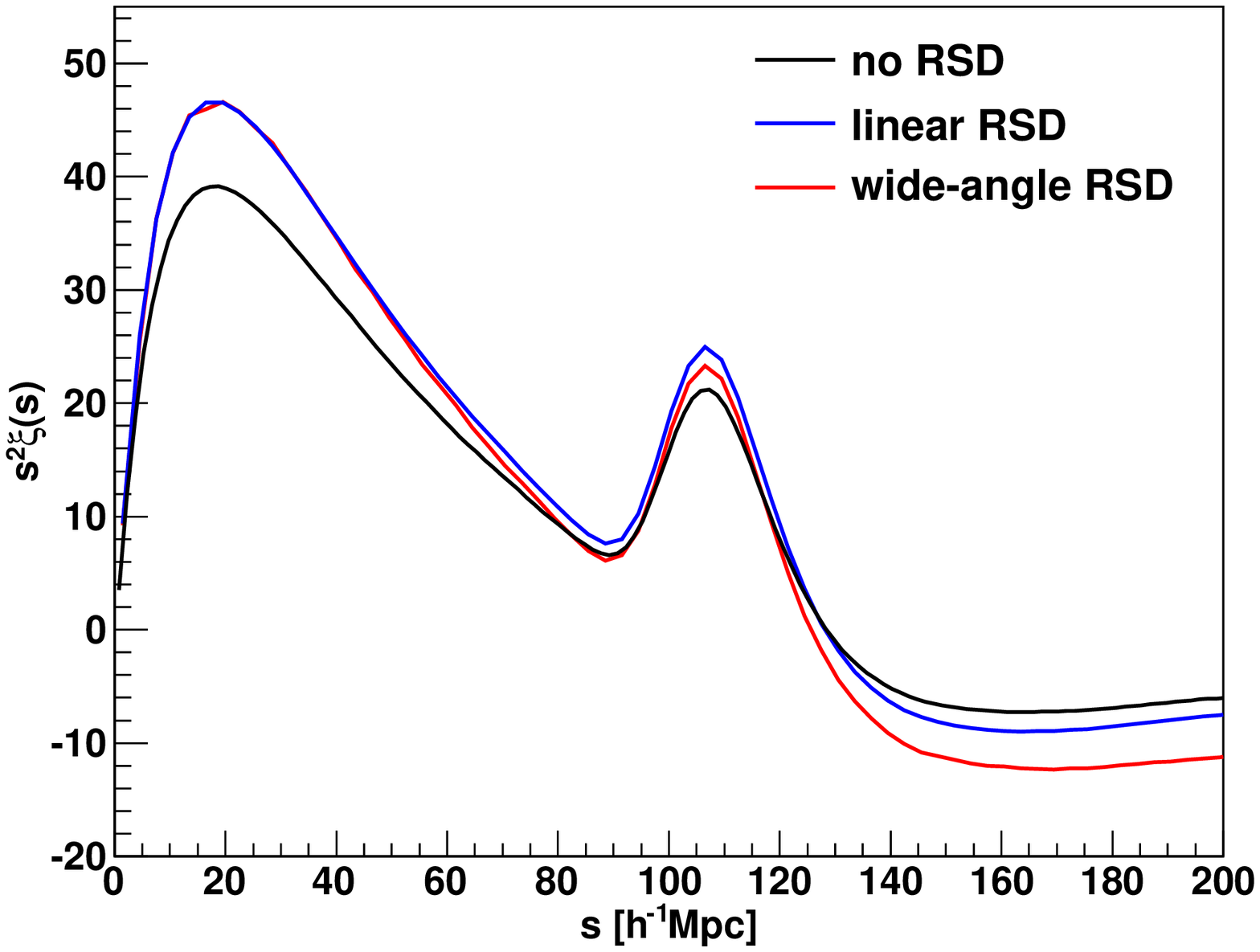,width=8cm}
\caption{The black line represents the plain correlation function without redshift space distortions (RSD), $\xi(r)$, obtained by a Hankel transform of our fiducial $\Lambda$CDM power spectrum. The blue line includes the linear model for redshift space distortions (linear Kaiser factor) using $\beta = 0.27$. The red line uses the same value of $\beta$ but includes all correction terms outlined in eq.~\ref{eq:wide1} using the $N(\phi,\theta,s)$ distribution of the weighted 6dFGS sample employed in this analysis.}
\label{fig:wide2}
\end{center}
\end{figure}

The final spherically averaged correlation function is given by
\begin{equation}
\xi(s) = \int^{\pi}_0\int_{0}^{\pi/2}\xi(\phi,\theta,s)N(\phi,\theta,s)\;d\theta d\phi,
\label{eq:mom}
\end{equation}
where the function $N(\phi,\theta,s)$ is obtained from the data. $N(\phi,\theta,s)$ counts the number of galaxy pairs at different $\phi$, $\theta$ and $s$ and includes the areal weighting $\sin(\phi)$ which usually has to be included in an integral over $\phi$. It is normalised such that 
\begin{equation}
\int^{\pi}_0\int^{\pi/2}_0 N(\phi,\theta,s)\;d\theta d\phi = 1 .
\end{equation}
If the angle $\theta$ is of order $1\;$rad, higher order terms become dominant and eq.~\ref{eq:wide1} is no longer sufficient. Our weighted sample has only small values of $\theta$, but growing with $s$ (see figure~\ref{fig:theta_s}). In our case the correction terms contribute only mildly at the BAO scale (red line in figure~\ref{fig:wide2}). However these corrections behave like a scale dependent bias and hence can introduce systematic errors if not modelled correctly. 

\label{lastpage}

\end{document}